\documentclass[3p]{earticle-nodate}

\PassOptionsToPackage{hyphens}{url}
\usepackage{graphicx}
\usepackage{dcolumn}
\usepackage{bm}
\usepackage[T1]{fontenc}
\usepackage[utf8]{inputenc}
\usepackage{amssymb}
\usepackage{amsmath}
\usepackage[dvipsnames]{xcolor}
\usepackage{placeins}
\usepackage{listings}
\usepackage{setspace}
\usepackage[normalem]{ulem}
\usepackage{url}
\usepackage{xr}
\usepackage{tabularx}
\usepackage{lmodern}

\biboptions{super}

\newcommand{\VEs}{\text{VEs}}
\newcommand{\VEd}{\text{VEd}}
\newcommand{\VEc}{\text{VEc}}
\newcommand{\VEi}{\text{VEi}}
\newcommand{\VE}{\text{VE}}

\begin{document}

\abstracttitle{Summary}

\begin{frontmatter}

\title{How will mass-vaccination change COVID-19 lockdown requirements in Australia?}

\author{Cameron Zachreson, PhD$^{1,2}$, Sheryl L. Chang, PhD$^{1}$, Oliver M. Cliff, PhD$^{1,3}$,  Mikhail Prokopenko, PhD$^{1,4}$}
\address{$^{1}$  Centre for Complex Systems, Faculty of Engineering, The University of Sydney, Sydney, NSW 2006, Australia\\
$^{2}$  School of Computing and Information Systems, The University of Melbourne, Parkville, VIC 3052, Australia\\
$^{3}$  Centre for Complex Systems, School of Physics, Faculty of Science, The University of Sydney, Sydney, NSW 2006, Australia\\
$^{4}$  Sydney Institute for Infectious Diseases, The University of Sydney, Westmead, NSW 2145, Australia\\ 
  \ \\
Correspondence to:\\ Prof Mikhail Prokopenko, Centre for Complex Systems\\ Faculty of Engineering, University of Sydney, Sydney, NSW 2006, Australia\\ \textbf{mikhail.prokopenko@sydney.edu.au} 
}


\begin{abstract}

\textbf{Background}

To prevent future outbreaks of COVID-19, Australia is pursuing a mass-vaccination approach in which a targeted group of the population comprising 
healthcare workers, aged-care residents and
other individuals 
at increased risk of exposure will receive a highly effective priority vaccine. 
The rest of the population will instead have access to a less effective vaccine.

\textbf{Methods}

We apply a large-scale agent-based model of COVID-19 in Australia to investigate the possible implications of this hybrid approach to mass-vaccination.  The model is calibrated to recent epidemiological and demographic data available in Australia, and accounts for several components of vaccine efficacy.

\textbf{Findings}

Within a feasible range of vaccine efficacy values, our model supports the assertion that complete herd immunity due to vaccination is not likely in the Australian context. For realistic scenarios in which herd immunity is not achieved, we simulate the effects of mass-vaccination on epidemic growth rate, and investigate the requirements of lockdown measures applied to curb subsequent outbreaks. In our simulations, Australia's vaccination strategy can feasibly reduce required lockdown intensity and initial epidemic growth rate by 43\% and 52\%, respectively.  
The severity of epidemics, as measured by the peak number of daily new cases, decreases by up to two orders of magnitude under plausible mass-vaccination and lockdown strategies.

\textbf{Interpretation}

The study presents a strong argument for a large-scale vaccination campaign in Australia, which  would substantially reduce both the intensity of future outbreaks and the stringency of non-pharmaceutical interventions required for their suppression.

\textbf{Funding}

Australian Research Council; National Health and Medical Research Council.

\end{abstract}

\end{frontmatter}

\clearpage

\section*{Introduction}

The Australian response to the COVID-19 pandemic has been very effective to date. 
Strict control measures, including travel restrictions and social distancing, successfully suppressed the initial pandemic wave in Australia (March -- June 2020)~\cite{chang2020modelling}, as well as several secondary outbreaks across the states, most notably in Victoria (June -- September 2020)~\cite{zachreson2021risk}.
However, as vaccines become available a more refined response is needed, given the need to balance population health against the high socio-economic impacts of local, regional and nation-wide lockdowns.  

The national COVID-19 vaccine rollout strategy developed by the Australian Government commenced in late February 2021, aiming to vaccinate a significant portion of the Australian population (the majority of the adult population) by the end of October 2021~\cite{hanly2021vaccinating}.
The first phase of the strategy targets priority groups with the BNT162b2 (Pfizer/BioNTech) vaccine, while the remainder of the population will receive the ChAdOx1 nCoV-19 (Oxford/AstraZeneca) vaccine during phases two and three. Both of these vaccines have demonstrated high clinical efficacy~\cite{dagan2021BNT162b2,knoll2021oxford}.  However, between August 2020 and January 2021, there has been a substantial increase in vaccine hesitancy in Australia, with 21.7\% of surveyed Australians responding that ``they probably or definitely would not get a safe and effective COVID-19 vaccine in January 2021''~\cite{biddle2021change}.
Furthermore, over April--June 2021, the  vaccine rollout strategy in Australia has been significantly revised due to health risks attributed to administering the ChAdOx1 nCoV-19 (Oxford/AstraZeneca) vaccine to individuals in specific age groups. As a result, from mid-June 2021 this vaccine is no longer recommended to Australians younger than 60, who instead became eligible for the BNT162b2 (Pfizer/BioNTech) vaccine.

Thus, many questions remain.
Is herd immunity achievable with current vaccination approaches?
To what extent can the strict lockdown rules be relaxed with a partial mass vaccination?
Is there an optimal but feasible balance between the vaccination efforts and social distancing practice?  What is the impact of the revised vaccine rollout strategy?
In this work, we approach these questions with a large-scale agent-based model (ABM) of COVID-19 transmission, case-targeted non-pharmaceutical interventions, lockdowns, and mass-vaccination in the context of Australia, using the latest available information.

There are several specific challenges in modelling COVID-19 vaccination campaigns: the complexity and burden of non-pharmaceutical interventions (NPIs); the heterogeneity of the population; country-specific demographics; logistical and supply constraints; as well as unknown vaccine characteristics.
The heterogeneity of the Australian population has been shown to unevenly affect the spread of respiratory diseases across different social contexts and wider jurisdictions~\cite{chang2020modelling,rockett2020revealing}. 
We can therefore expect complex trade-offs between NPIs and vaccination interventions, covering overlapping but not identical parts of the population.
These effects may be difficult to predict for situations in which the vaccine efficacy differs with respect to reducing susceptibility, preventing symptoms of infection, and limiting further transmission of the virus.
Some of the available vaccines, most notably BNT162b2 (Pfizer/BioNTech), have shown a high efficacy against documented infection, as well as symptomatic and severe disease~\cite{dagan2021BNT162b2}.
However, comprehensive results across multiple efficacy components are still lacking. 
In this work, we account for differences in vaccine efficacy for the two distinct vaccine types approved for distribution in Australia: a {\textit{priority}} vaccine, (e.g.., BNT162b2), and a {\textit{general}} vaccine, (e.g., ChAdOx1 nCoV-19).

In order to capture population heterogeneity, we adapted a previously developed and validated high-resolution ABM of mitigation and control of the COVID-19 pandemic in Australia~\cite{chang2020modelling,rockett2020revealing}.
This model included a range of dynamically adjustable NPIs, such as travel restrictions, case isolation, home quarantine and mandated social distancing (lockdown).
Here, we extended the ABM to include several detailed vaccination measures.
These extensions included an explicit account of separate components of vaccination efficacy (susceptibility, disease, and infectiousness), and changeable levels of age-stratified mass-vaccination coverage with the general and priority vaccines.

This paper addresses several open questions surrounding vaccination in Australia.
Firstly, we investigate the feasibility of herd immunity following a mass-vaccination campaign.
This is addressed by varying vaccination coverage with different vaccine efficacy combinations.
Secondly, we quantify the benefit of the general vaccine in scenarios where all priority vaccine supplies are consumed by considering different levels of general vaccination distributed in addition to a fixed realistic priority vaccination coverage.
Finally, we quantify to what extent mass-vaccination can reduce or eliminate the need for lockdowns by varying lockdown compliance levels for various extents of vaccination coverage.

\section*{Methods}

\subsection*{Simulating COVID-19 in Australia}

Our approach to simulating COVID-19 in Australia follows that of our previous work \cite{chang2020modelling}.  This model is implemented within a high-precision simulator comprising about
23.4 million stochastically generated software agents. These artificial ``agents'' represent the population in Australia, with attributes of an anonymous individual (e.g., age, residence, gender, workplace, susceptibility and immunity to diseases), and contact rates within different social contexts (e.g., households, household clusters, neighbourhoods, classrooms, workplaces). The set
of agents, i.e., the surrogate population, is generated to match the average characteristics of the Australian Census and the Australian Curriculum, Assessment and Reporting Authority data, including commuting patterns between the places of residence (i.e., census statistical areas) and work or study (i.e., census destination zones)~\cite{cliff2018nvestigating,zachreson2018urbanization,fair2019creating}. Furthermore, the model is calibrated to key COVID-19 characteristics, using age-dependent contact and transmission rates (scaled to match the COVID-19 reproductive number $R_0$ by the scaling factor $\kappa$), the fraction of symptomatic cases (set as 0.134 for children, and 0.669 for adults), the  probabilities of transmission for asymptomatic/presymptomatic and symptomatic agents, and other parameters specifying the natural disease history model (see below)~\cite{chang2020modelling}.

A discrete-time simulation scenario progresses by updating agents’ states over time, starting from an initial distribution of infection, seeded by imported cases dependent on the incoming international air traffic (using data from the Australian Bureau of Infrastructure, Transport and Regional Economics)~\cite{cliff2018nvestigating,zachreson2018urbanization}.  Unless travel restrictions (i.e., border closures) are imposed by a scenario, at each time step this process probabilistically introduces new infections within a 50 km radius of every international airport, in proportion to the average daily number of incoming passengers at that airport (using a binomial distribution)~\cite{cliff2018nvestigating,zachreson2018urbanization,chang2020modelling}.
By simulating interactions within all the mixing  contexts in the surrogate population in 12-hour cycles (``day'' and ``night''), with respect to work, study and other activities, a specific outbreak, originated at particular points, is traced over time.  A typical distribution of local transmissions (distinct from overseas acquired cases), traced from the simulation, has been cross-validated with the genomic surveillance data in Australia~\cite{rockett2020revealing}.

The following modifications were made to our model of COVID-19 disease natural history and case ascertainment: 

\begin{itemize}
 \item {Infectious incubation times ($T_{\text{inc}}$) calibrated to the findings of Ferretti et al. {\cite{ferretti2020quantifying}}, and Lauer et al. \cite{lauer2020incubation}, who inferred log-normally distributed incubation times with means of approx. 5.5 days.} 
    \item{An infectious asymptomatic or symptomatic period ($T_\text{symp}$), following incubation, lasting between 7 days and 14 days (uniformly distributed), based on estimates of the replication-competent viral shedding period used to support guidance on case isolation periods published by the United States Centre for Disease Control and Prevention (CDC, see~\cite{cdc2021interim} and references therein \cite{wolfel2020virological}).} 
    \item{Differentiation between ``asymptomatic infectivity" and ``pre-symptomatic infectivity". In the previous iteration of the model used in Chang et al.~\cite{chang2020modelling}, asymptomatic and pre-symptomatic individuals had reduced infectivity to contacts. That assumption was modified in this work, for which pre-symptomatic cases are assumed to be as infectious as symptomatic cases (with respect to viral load), while those who remain asymptomatic throughout the course of disease have reduced infectivity (a factor of 0.5 is applied to the force of infection exerted on contacts). This change reflects the general finding that pre-symptomatic transmission is responsible for a substantial amount of COVID-19 spread (up to 50\% of transmission), and allows a parsimonious calibration of disease natural history, reproductive ratio, and generation interval \cite{ferretti2020quantifying,wu2020estimating}.}
    \item{To simulate the imperfect detection of cases in a scenario with high levels of voluntary population screening, we introduce two case detection probabilities, one for symptomatic case detection and the other for pre-symptomatic and asymptomatic detection. For symptomatic cases, the probability of detection per day is set to  0.23, while for pre-symptomatic and asymptomatic cases, the probability of detection per day is 0.01.} 
\end{itemize}

Interventions are specified via suitably defined macro- and micro-parameters constraining agent interactions and transmission probabilities. These constraints represent assumptions on how non-pharmaceutical (e.g., social distancing)  or pharmaceutical (e.g., vaccine efficacy) interventions reduce the infection spread. For example, social distancing compliance can be set at 80\% at the macro-level, while micro-distancing contacts during a lockdown can be reduced to 10\% within workplaces and 25\% within communities (cf. Table S3).

\subsection*{Model calibration}
Because the models of disease natural history and case ascertainment were modified, we re-calibrated the model used by Chang et al.~\cite{chang2020modelling} to approximately match the case incidence data recorded during the first and second waves of COVID-19 in Australia and the global reproduction number of $R_0 \approx 2.9~95\%~\text{CI}~[2.39, 3.44]$~\cite{billah2020reproductive} (see Supplementary Material).

\subsection*{Mass-vaccination simulations}

In our mass-vaccination scenarios, we used an age-stratified vaccine allocation scheme. Starting with no individuals vaccinated, the algorithm allocates new immunisations randomly according to the following ratio: 100:10:1, which correspond to [age $\geq 65$] : [$18 \leq \text{age} < 65$] : [age $< 18$ ]. That is, for every 100 individuals aged over 64 years, 10 individuals aged between 18 and 64 years are immunised and one individual under the age of 18 years is immunised. This allocation ratio applies unless there are no remaining unvaccinated individuals in an age category, in which case vaccines are allocated to the remaining age categories according to the same specified proportions until the specified number of immunisations is depleted or all individuals over the age of 18 are immunised, whichever occurs first. The priority vaccines are distributed first, followed by the general vaccines. In practice, due to the age distribution in our model of the Australian population (based on the 2016 ABS Census) this means that all individuals aged over 64 years (i.e., 65+) are immunised unless there are fewer than $3.9 \times 10^6$ total immunisations. In our hybrid mass-vaccination scenarios, we assume at least $5 \times 10^6$ priority immunisations, so the entire population over the age of 64 years is immunised with the priority vaccine, while the remaining immunisations are distributed between children and adults under 65 in a ratio of 10/1 (adults/children). All immunisations are allocated on day 0 of each outbreak simulation.  Note that because our allocation procedure terminates after all individuals aged 18 or older are immunised, the maximum number of vaccinated children is 1/10 of the population between 18 and 64, or 1.4M children vaccinated. This accounts for approximately 25\% of the population under 18, reflecting the tighter regulations on vaccine approval for these age groups worldwide, which are currently relaxing for adolescents over the age of 12. Therefore, the maximum number of immunisations is capped at 19.3M, leaving approximately 4.2M children unvaccinated.

\subsection*{Growth rate estimation}

Throughout this work, we report growth rates estimated from incidence data produce by the ABM or collected from government case reports. To estimate growth rates, we fit each case incidence timeseries to a delayed exponential function: 
\begin{equation}
    I(t) = \exp(\lambda (t - \Delta t))\,,
\end{equation}
where $\lambda$ is the exponential growth rate of case incidence, and the delay $\Delta t$ accounts for transient stochastic effects during the early stages of outbreaks as well as delays in detection of new cases.

\subsection*{Role of the funding source}
The funders of the study had no role in the study design; the collection, analysis, and interpretation  of  data; the  writing of the Article; and in the decision to submit the paper for publication. The corresponding author had full access to all the data in the study and had final responsibility for the decision to submit for publication.

\section*{Results}

We present our results in three sections, first covering questions related to herd immunity and vaccine efficacy, then the effect of mass-vaccination on epidemic growth rate, and lastly the effects of  future lockdowns in conjunction with  prepandemic mass-vaccination.
Section~A  
investigates the feasibility of achieving herd immunity given the existing known and unknown aspects of COVID-19 vaccine efficacy.
For clinical efficacy values of 0.9 and 0.6 (corresponding to  conservative estimates for the priority and the general vaccine, respectively), we use a homogeneous approximation to calculate the coverage required to achieve herd immunity as a function of vaccine efficacy against susceptibility, symptom expression, and onward transmission.
We then compare the results of our ABM to the homogeneous approximation for a subset of vaccine efficacy values.
Section~B  
describes the effects of realistic simulated mass vaccination regimes on epidemic growth rate.
In scenarios where herd immunity is not achieved, we compute the growth rate of cumulative incidence for different levels of vaccination coverage, in combination with case-targeted NPIs.
Section~C  
investigates outbreak suppression in mass-vaccination scenarios and shows how vaccination can reduce the fraction of the population required to be in lockdown to achieve suppression of case incidence.

\subsection*{A. Herd immunity requirements: coverage and efficacy}\label{results_A}
For the purposes of modelling the effects of vaccination on the spread of COVID-19 and evaluating the requirements for herd immunity, three main components of vaccine efficacy constitute important unknown factors:
\begin{itemize}
    \item Efficacy for susceptibility ($\VEs$) determines the level of immunity vaccination imparts to those susceptible to the virus. In the ABM this parameter reduces the probability of becoming infected if exposed.
    \item Efficacy for disease ($\VEd$) determines the expression of illness in those who are vaccinated and subsequently become infected. In the ABM this parameter reduces the probability of expressing symptoms if infected.  
    \item Efficacy for infectiousness ($\VEi$) reduces the potential for vaccinated individuals to transmit the virus if infected. In the ABM, this parameter reduces the force of infection produced by infected individuals who are vaccinated. 
\end{itemize}
The practical bounds of the efficacy terms ($\VEs$, $\VEi$, and $\VEd$) are only partially constrained by the clinical efficacy ($\VEc$) reported in clinical trials, since:
\begin{equation}\label{Eq_VEc}
    \VEc = \VEd + \VEs - \VEs \ \VEd.
\end{equation}
The clinical efficacy, $\VEc$, is defined as the reduction in presentation of clinical disease in the vaccine group relative to the control.
The vaccine efficacy for an individual's susceptibility and disease, $\VEs$ and $\VEd$, are constrained by the clinical efficacy, $\VEc$, through Eq.~\eqref{Eq_VEc}; however, the vaccine efficacy for infectiousness, $\VEi$, is left undefined by clinical trial data.
Although unreported, it is crucial for $\VEi$ to be defined in order to compute the population-level vaccine efficacy, $\VE$, for a given proportion of the population vaccinated.
For defined values of the input parameters (coverage, $\VEi$, $\VEs$, and $\VEd$), $\VE$ can be estimated from a homogeneous approximation of population mixing (see Supplementary Material) and the necessary vaccine coverage threshold for herd immunity can then be estimated by computing the effective reproductive number, $R$, after vaccination: 
\begin{equation}
    R = R_0 (1 - \VE),
\end{equation}
where $R_0$ is the basic reproductive ratio in a completely susceptible population, and herd immunity is achieved when $R < 1$.
In the Supplementary Material Fig.~S1, 
we provide herd immunity thresholds in the $\VEi\times \text{coverage}$ plane for different combinations of $\VEd$ and $\VEs$ constrained through Eq.~\eqref{Eq_VEc}, with $R_0 = 2.75$ to match the basic reproductive ratio used in our ABM.

These results show that, for herd immunity in a homogeneous system, the general vaccine (with $\VEc = 0.6$) must produce a substantial reduction in the transmission potential of infected individuals, with an efficacy for infectiousness ($\VEi$) on the same order of efficacy for susceptibility ($\VEs$) and disease ($\VEd$). On the other hand, Fig.~S1b 
illustrates that a priority vaccine with clinical efficacy of $90\%$ ($\VEc = 0.9$) could produce herd immunity with vaccination coverage between $64\%$ and $86\%$. Moreover, as long as $\VEs \geq \VEd$, this result can be achieved without a substantial contribution from the unknown effect of the vaccine on reducing infectiousness (i.e., $\VEi$).

Of course, Australia is not a homogeneous system with respect to population mixing patterns, and the true vaccine allocation strategy is intentionally heterogeneous, first prioritising those in high-risk age groups and prioritising children last.
For these reasons, we expected our ABM to produce results differing from those estimated by the homogeneous approximation.
To match our scenarios to the Australian context, which has consistently maintained case-targeted non-pharmaceutical interventions, we performed a systematic scan of efficacy parameters both with and without combinations of detected case isolation, home quarantine of household contacts, and international travel restrictions.
While the proposed vaccination regime in Australia involves both priority and general vaccines,  
initially we treat each separately in order to reduce the complexity of the parameter space.

Therefore, the results presented in this section are not directly applicable to proposed vaccination levels in Australia, but can be used to guide intuition with respect to the influence of different efficacy combinations and potential deviation from estimates based on homogeneous approximations. 

For a given mass-vaccination scenario, the ABM mimics the current government roll-out policy by allocating immunisations using an age-stratified system.
In this system, individuals aged 65 and older are preferentially vaccinated, with second preference for those aged 18 - 64 years, and third preference for those under the age of 18 (see Methods). We do not explicitly simulate partial vaccination (i.e., using one dose only), with the number of individuals immunised corresponding to the prospective number of completed vaccine treatment schedules.  In this analysis we relax the condition that limits vaccine allocation to only 25\% of the child population, instead limiting allocation only by the number of vaccines distributed. We made this choice to ensure that our parameter sweeps were not bounded by the limitations currently in place on vaccination of children, allowing a more complete scan of the parameter space available to the model and an effective identification of herd immunity thresholds. 

\begin{figure}
    \centering
    \includegraphics[width=\textwidth]{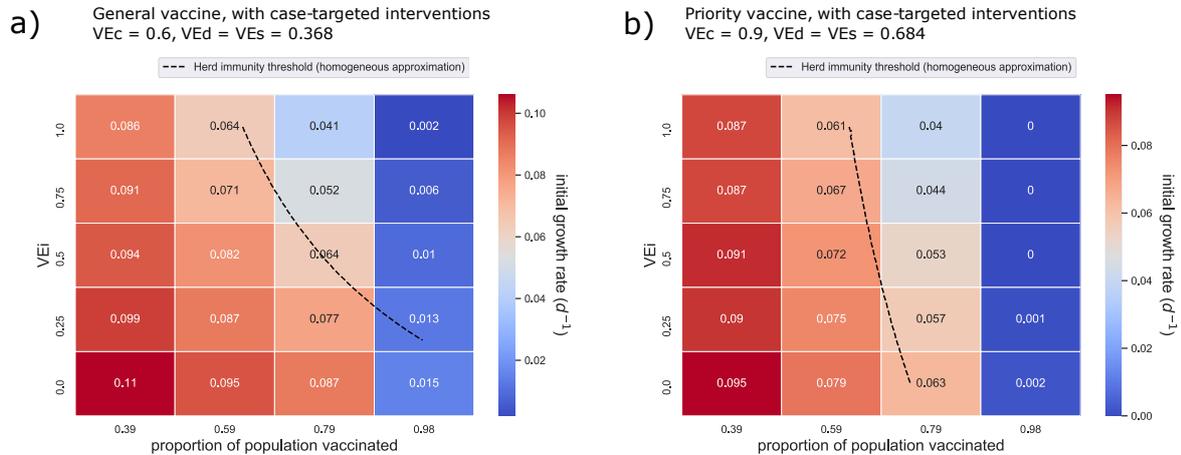}
    \caption{Simulations suggest that herd immunity is unlikely to be attained by either the general or the priority vaccine alone. The incidence growth rates shown here computed from results of the ABM over a range of values for coverage and vaccine efficacy against infectiousness ($\VEi$) for the general vaccine (a, $\VEc = 0.6$) and the priority vaccine (b, $\VEc = 0.9$). Here, central values of efficacy against disease and susceptibility were used ($\VEd = \VEs = 1 - \sqrt{1 - \VEc}$), and case-targeted NPIs were applied in addition to vaccination.}
    \label{fig:coverage_threshold_ABM_CTNPI_gr}
\end{figure}

Our full results are given in the Supplementary Material Tables S5 
through S16, 
and are summarised here below. Note that in the ABM, due to continuous introductions of cases from overseas, the clustering of children into school contact networks, and age-stratified vaccine roll-out, the vaccine coverage required for herd immunity is substantially higher than predicted by the homogeneous approximation  (Fig.~\ref{fig:coverage_threshold_ABM_CTNPI_gr}).

Our results show that the general vaccine ($\VEc = 0.6$) alone cannot feasibly induce herd immunity, even when combined with targeted non-pharmaceutical interventions. Specifically, the general vaccine by itself (without case-targeted NPIs) can only produce herd immunity if efficacy for infectiousness (VEi) is above 0.5 (i.e., on the order of VEd and VEs, see Fig.~S2a,
with a coverage threshold between $80\%$ and $98\%$ corresponding to a nonlinear drop in epidemic growth rate, after which growth rapidly approaches zero. This result suggests that even if all individuals over 18 were vaccinated, and VEi were sufficiently high, a portion of the child population would need to be immunised in order to achieve herd immunity. However, there are still significant benefits for feasible ranges of vaccine coverage and efficacy.

Notably, peak prevalence can be reduced by a factor of two without the need for case-targeted interventions, with central values of $\VEd$ and $\VEs$ (i.e., when $\VEd = \VEs = 1 - \sqrt{1 - \VEc} = 0.368$), a reasonable value of $\VEi \approx 0.5$, and only $40\%$ population coverage (Tab.~S6).
Referring to Fig.~S3a,
increasing this vaccine coverage to approximately $80\%$ reduces peak prevalence by $83$--$88\%$ 
(Tab.~S6)
and delays the peak by several weeks 
(Tab.~S7).
By combining case-targeted interventions with this $80\%$ coverage, the general vaccine can dramatically slow the spread of the virus. For VEi = 0.5, growth rate is reduced by 53\% from the baseline value of 0.137$d^{-1}$ for unmitigated outbreaks down to 0.064$d^{-1}$ (Fig.~\ref{fig:coverage_threshold_ABM_CTNPI_gr}a, Tab.~\ref{tab:growth_rate}). Even when herd immunity is not reached, this reduction in epidemic growth rate corresponds to large reductions in epidemic intensity as measured by the peak number of concurrent active cases 
(Fig.~S4a).

The priority vaccine ($\VEc = 0.9$), produces similar benefits with a vaccine coverage as low as 60\% and central efficacy values of $\VEs = \VEd = 0.684$, $\VEi \in [0.5, 0.75]$ 
(Figures~S2b~and~S3b).
In combination with targeted interventions, our simulations suggest that the priority vaccine could produce herd immunity, with a coverage threshold between $80\%$ and $98\%$. This threshold exists even if  efficacy against infectiousness is negligible (VEi $\approx$ 0), with the coverage required for herd immunity decreasing gradually with higher VEi (Fig.~\ref{fig:coverage_threshold_ABM_CTNPI_gr}b). However, the current  vaccination rollout in Australia constrains the maximum coverage of the priority vaccine, relying on a combination of priority and general vaccines. Therefore, substantial uptake of the less-effective general vaccine will be required if the benefits of the priority vaccine are to be realised for the whole population. For a more feasible coverage of $40\%$, the priority vaccine combined with case targeted NPIs reduces the epidemic growth rate by $34\%$ (with VEi = 0.5) and delays the peak by approximately 40 days (Fig.~\ref{fig:coverage_threshold_ABM_CTNPI_gr}b, 
Fig.~S4b, Tab.~S16).
As a benchmark value, note that our simulations show case-targeted interventions alone decrease growth rate by 14\% relative to unmitigated outbreaks (Tab. \ref{tab:growth_rate}). 

To summarise, herd immunity is possible in principle for both general and priority vaccines, however, it requires coverage levels above 80\% (of the entire population) regardless of whether or not case-targeted NPIs are implemented. Furthermore, for the general vaccine, herd immunity is not achievable at any coverage for VEi < 0.5, which is an optimistic upper bound for the general vaccine~\cite{harris2021impact}. The results of the ABM are {\it qualitatively} consistent with the homogeneous approximation, but indicate that heterogeneity produces substantial {\it quantitative} differences in the coverage levels required for herd immunity. These levels are higher than what is currently achievable under the existing mass-vaccination strategy in Australia. Assuming limited vaccination in the Australian population younger than 18, overall vaccination coverage is constrained to less than 80\%, even if the entire adult population is vaccinated. 

\FloatBarrier

\subsection*{B. Effects of hybrid mass-vaccination on epidemic growth}\label{results_B}

To investigate the possible effects of realistic mass-vaccination strategies on epidemic growth dynamics, we selected central values of efficacy for the priority vaccine ($\VEi = \VEs = \VEd = 0.684$) and general vaccine ($\VEi = \VEs = \VEd = 0.368$), and simulated initial epidemic growth in eight different scenarios (Tab.~\ref{tab:scenarios}). We find that realistic hybrid vaccination campaigns systematically reduce epidemic growth rate with increasing coverage (by up to a factor of two). However, we do not identify a distinct coverage level beyond which growth rate decreases sharply (i.e., a herd immunity threshold). 

To improve the realism of our model for simulating hybrid vaccination scenarios, we use the latest estimates of coverage with the priority vaccine (enough has been {purchased at the time of writing to vaccinate up to 10M individuals, or approximately $40\%$ of the population), and the general vaccine.
We assume that the general vaccine will not be subject to supply constraints, with coverage limited instead by uptake.
To reflect the current situation in Australia where the disease is currently controlled,
we do not simulate a progressive vaccine rollout during the outbreak.
 We made the decision not to simulate progressive rollout for two reasons:
\begin{enumerate}
    \item The timescale of COVID-19 outbreaks and the associated policy response has so far been much faster than the timescale of vaccination (substantial changes in vaccination levels require weeks or months, while outbreaks have typically triggered lockdown implementation within days). 
    \item{Simulating progressive rollout requires several additional degrees of freedom, so that capturing the details of progressive rollout increases the complexity of the model without substantially improving the insight available from the results it generates.}
\end{enumerate}
Instead, we explore different coverage levels as a proxy for timing of the next epidemic wave between the beginning and end of the protracted vaccination campaign.
We compare scenarios ranging from a relatively small number of priority vaccine immunisations (5M two-dose vaccinations), to an optimistic endpoint scenario with 10M priority (two-dose) immunisations and an additional 9.3M general immunisations, for coverage of 82\% (100\% of the adult population, and 25\% of the population aged less than 18 years).

  \setlength{\arrayrulewidth}{0.5mm}
    \setlength{\tabcolsep}{18pt}
    \renewcommand{\arraystretch}{1.5}
    \begin{table}
    \centering
    \resizebox{\textwidth}{!}{
    \begin{tabular}{c| c | c | c }
        scenario & targeted NPIs & priority immunisations & general immunisations    \\
        \hline \hline
         no intervention & nil & nil & nil \\\hline
         targeted NPIs only & TR, CI, HQ*  & nil  & nil \\\hline
         priority vaccine (5M) & ''  &  $5\times10^6$ & nil \\\hline
         general vaccine (11.5M) & '' & nil & $11.5\times10^6$  \\ \hline
         priority vaccine (10M) & ''  &  $10^7$  & nil \\\hline
         priority 10M, general 2.5M  & '' &  $10^7$ & $2.5\times10^6$ \\\hline 
         priority 10M, general 6.1M  & '' & $10^7$ & $6.1\times10^6$ \\ \hline
         priority 10M, general 9.3M  & '' & $10^7$ & $9.3\times10^6$ \\ \hline
    \end{tabular}}
    \caption{Selected vaccination scenarios simulated with the ABM. The numbers under ``priority immunisations" and ``general immunisations" correspond to the number of individuals who have undergone a full vaccination regime (i.e., a two-dose regime for the priority vaccine). *TR: travel restrictions (ban on international travel), CI: case isolation (in-home isolation of detected cases), HQ: home quarantine (in-home isolation for household contacts of detected cases).}
    \label{tab:scenarios}
    \end{table}

For each scenario, we simulated 110 realisations of outbreaks and estimated the expected growth rate of cumulative incidence for the first 2\,000 cases (Fig.~\ref{fig:growth_rate}). To do so, we computed the incidence growth rate of each realisation (see Methods) and estimated the mean growth rate over all realisations of each mass-vaccination scenario (Tab.~\ref{tab:growth_rate}). Initial growth of case incidence decreases gradually as vaccination levels increase (Fig.~\ref{fig:growth_rate}). With 10M priority immunisations (maximum projected supply), the growth rate decreased by 28\% relative to the rate computed with targeted NPIs only. At a feasible endpoint condition with $82\%$ of the population vaccinated (10M priority and 9.3M general immunisations), the average growth rate decreases by $52\%$ (from 0.118$~d^{-1}$ with targeted NPIs only, to 0.057$~d^{-1}$ with vaccination). In this scenario, the lockdown compliance required for epidemic suppression decreased by 43\% (from $\approx 70\%$ with targeted NPIs only to $\approx 40\%$ with 82\% vaccination coverage).  

Log-scaled plots of initial case incidence (Fig.~\ref{fig:growth_rate}a) clearly demonstrate the lack of a defined herd immunity threshold within the set of plausible scenarios we investigated. This is not surprising given that the levels of vaccine coverage reached are lower than those required for herd immunity with either vaccine individually (Fig \ref{fig:coverage_threshold_ABM_CTNPI_gr}). The distribution of the first 2000 cases between age groups suggests that systematically placing children at low priority for immunisation increases the required threshold for nonlinear reductions in epidemic growth rate (Fig.~\ref{fig:growth_rate}b),  consistent with our results for single vaccine types (Fig.~\ref{fig:coverage_threshold_ABM_CTNPI_gr}). Based on the risk-averse precedent for COVID-19 response in Australia, it is plausible that even the relatively slow spreading rates we calculate for the endpoint vaccination scenario would lead to some level of lockdown imposition, which we address in the following section.

\begin{figure}
    \centering
    \includegraphics[width=0.7\textwidth]{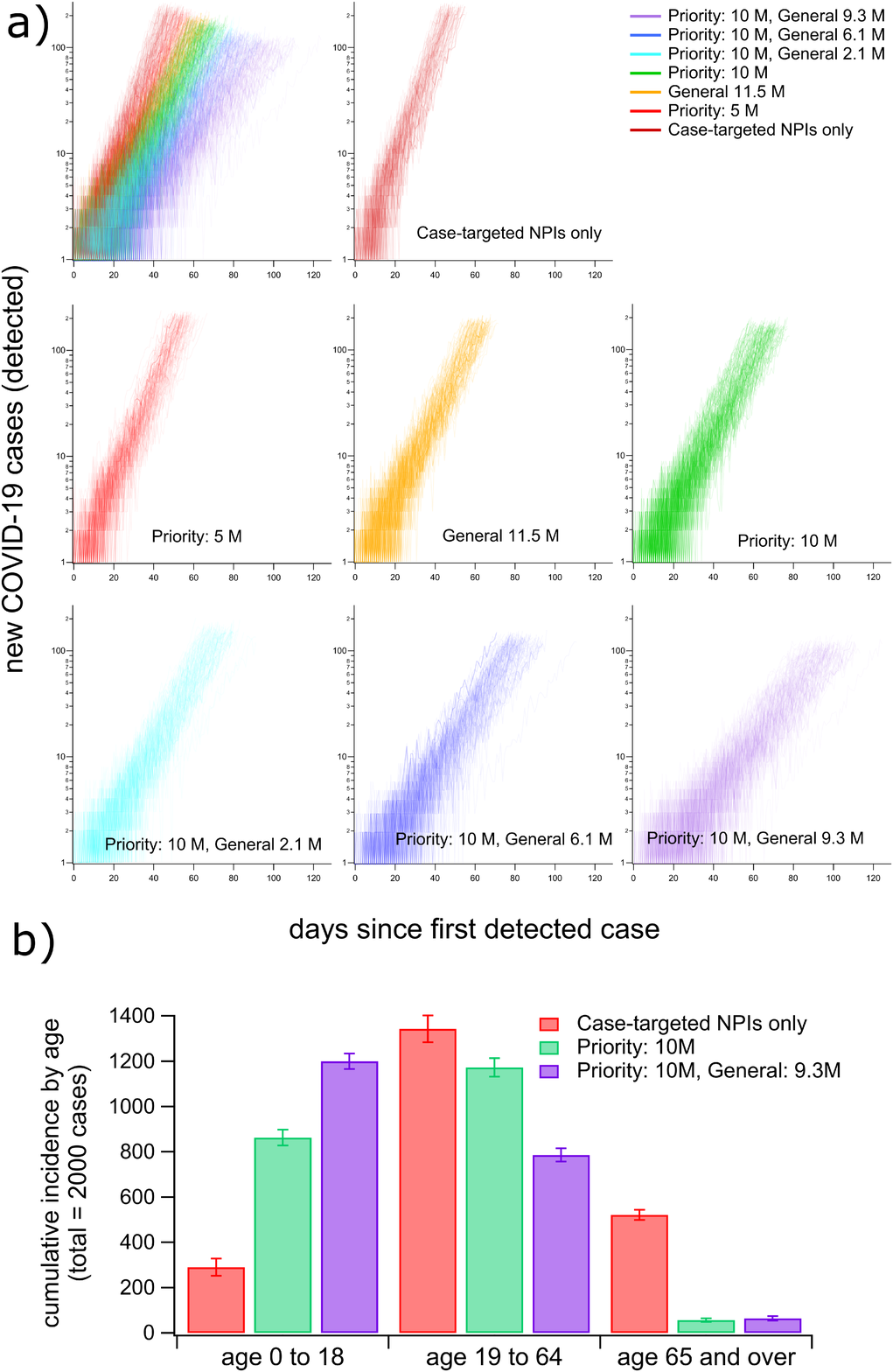}
    \caption{Hybrid vaccination programs produce up to a two-fold reduction of the epidemic growth rate. Individual incidence trajectories are colour-coded by mass-vaccination scenario (110 trajectories are shown for each). Each trajectory ends at the time the lockdown trigger condition was reached (cumulative incidence exceeding 2000 cases). The plots in (a) show log-scaled incidence trajectories for each vaccination scenario. Subplot (b) shows the distribution of the first 2000 cases in the three age groups used to prioritise vaccination in three representative scenarios (error bars show standard deviations over 110 realisations). Summary growth rate statistics for each scenario are given in Tab.~\ref{tab:growth_rate}.}
    \label{fig:growth_rate}
\end{figure}

\setlength{\arrayrulewidth}{0.5mm}
\setlength{\tabcolsep}{18pt}
\renewcommand{\arraystretch}{1.5}
\begin{table}
    \centering
    \resizebox{\textwidth}{!}{
    \begin{tabular}{c|c|c|c}
        scenario & mean growth rate & quantiles [5\%, 95\%]  & 95\% CI (mean, bootstrap) \\
        \hline \hline
         no intervention & 0.137 & [0.128, 0.146] & [0.1356, 0.1376]
         \\\hline
         targeted NPIs only & 0.118  & [0.110, 0.127]  & [0.1171, 0.1189 ] \\\hline
         priority vaccine (5M) & 0.104  & [0.097, 0.112]  & [0.1034, 0.1052] \\\hline
         general vaccine (11.5M) & 0.091 & [0.083 0.099] & [0.0901, 0.0917]  \\ \hline
         priority vaccine (10M) & 0.085  & [0.076, 0.092]  & [0.0843, 0.0859] \\\hline
         priority 10M, general 2.5M & 0.078 & [0.072, 0.087] & [0.0771, 0.0787] \\\hline 
         priority 10M, general 6.1M  & 0.067 & [0.061, 0.072] & [0.0666, 0.0679] \\ \hline
         priority 10M, general 9.3M  & 0.057 & [0.052, 0.062] & [0.0562, 0.0573] \\ \hline
    \end{tabular}}
    \caption{Growth rates of daily incidence for eight different intervention scenarios. For each scenario, growth rates were computed for 110 realisations. The values shown here are ensemble means from each scenario, as well as the 5\% and 95\% quantiles of the growth rate distribution from each set of realisations and the 95\% bootstrap confidence interval for the mean.}
    \label{tab:growth_rate}
\end{table}

\FloatBarrier

\subsection*{C. Effects of mass-vaccination on lockdown requirements}\label{results_C}

For the realistic vaccination scenarios given in Tab.~\ref{tab:scenarios}, we estimated the level of lockdown compliance required to suppress epidemic growth.
In each scenario, the population-scale physical distancing (lockdown) measures were enacted when the epidemic reached cumulative incidence of 2\,000 cases.  Choosing this relatively high threshold for the implementation of lockdown allows us to illustrate how epidemic dynamics depend on vaccination levels, and how these dynamics respond to the implementation of lockdown restrictions (Fig.~\ref{fig:TS_hybrid}). Incidence initially increases exponentially, and the growth rate is reduced after the imposition of lockdown restrictions (occurring at approximately day 50 in Fig.~\ref{fig:TS_hybrid}a, and at day 120 in Fig.~\ref{fig:TS_hybrid}b). If enough of the population complies with physical distancing measures, the growth rate becomes negative and the conditions for eventual suppression are met. 
In concordance with our previous work~\cite{chang2020modelling}, this threshold lies between 60\% and 70\% compliance when only NPIs are considered.
In these simulations, vaccination systematically decreased the fraction of the population required to maintain physical distancing restrictions in order to suppress epidemic spread.
Summary results shown in Fig.~\ref{fig:severity_SD} demonstrate how this compliance threshold depends on the level of vaccination. At the endpoint condition of 10M priority immunisations and 9.3M general immunisations, the lockdown compliance threshold drops to approximately 40\% (Figures~\ref{fig:TS_hybrid}b~and~\ref{fig:severity_SD}).  

In the Supplementary Material, we show results obtained for higher general vaccine efficacy (raised from VEc = 0.6 to VEc = 0.75, Fig.~S7a), 
demonstrating that this approximate threshold was not sensitive  within this range to the precise efficacy value used for the general vaccine. In addition, we carried out sensitivity analysis in terms of efficacy against infectiousness for the priority vaccine, contrasting higher (VEi = 0.684) and lower levels (VEi = 0.5). The corresponding results, summarised in Supplementary Fig.~S7b, 
show that the reported outcomes are robust to this change as well.

Finally, we examined the sensitivity of our results to the priority structure of vaccine rollout (Figures S7b~and~S7c). 
Specifically, we re-parameterised the priority schedule to invert the prioritisation of individuals aged 65 or older relative to those aged 18-64, while holding constant the total vaccinated proportion in each age group. The requirements of lockdown under this inverted priority structure are practically invariant (Fig.~S8).
This result reflects a low sensitivity of the model to priority levels when overall vaccination coverage remains unchanged.

In addition to reducing the lockdown compliance threshold, vaccination reduces the growth rate of the epidemic even in the absence of lockdown restrictions, so incidence levels during  outbreaks decrease dramatically for the same proportion of the population complying with physical distancing. For example, Fig.~\ref{fig:severity_SD} demonstrates that, 60 days after the beginning of lockdown, the complete vaccination program consistently decreases case incidence by almost two orders of magnitude, independently to the proportion of the population complying. Such a difference in case incidence could be expected to have a dramatic impact on the projected strain to medical infrastructure, even without considering vaccine efficacy for severe disease. Additionally, Fig.~\ref{fig:growth_rate} and Fig.~\ref{fig:severity_SD} demonstrate that, due to slower epidemic growth with vaccination, case incidence at the onset of lockdown decreases from approximately 250 cases per day (targeted NPIs only), down to 100 cases per day (completed vaccination program). Taken together, these results indicate that vaccination would allow for shorter, less restrictive physical distancing mandates, if such measures were required in order to suppress subsequent outbreaks.

\begin{figure}
    \centering
    \includegraphics[width = \textwidth]{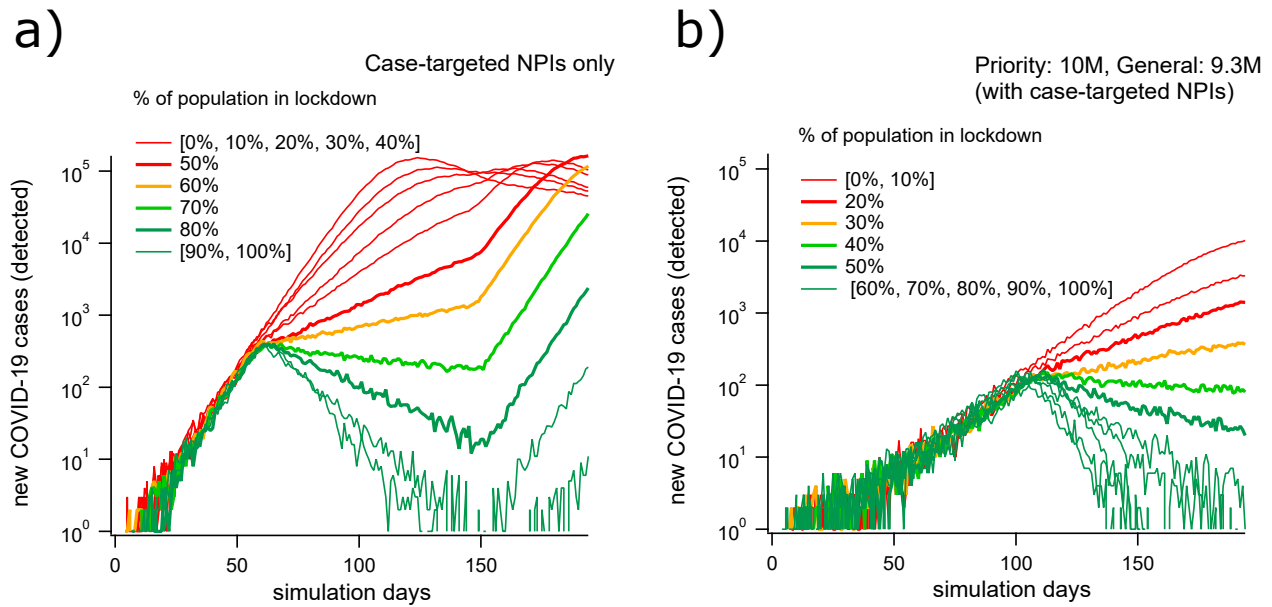}
    \caption{Our simulations suggest that realistic hybrid vaccinations strategies reduce the required intensity of lockdowns (mandated physical distancing) by a factor of two. Timeseries plots of representative case incidence trajectories for the scenario with targeted NPIs only (a) demonstrate a lockdown compliance threshold for elimination lying between 60\% and 70\%. Similar plots for the vaccination scenario with 10M priority vaccinations (VEc = 0.9; VEi = VEs = VEd = 0.684) and 9.3M general vaccinations (VEc = 0.6; VEi = VEs = VEd = 0.368), in addition to case-targeted NPIs, (b) show a lockdown compliance threshold for elimination lying between 30\% and 40\%.}
    \label{fig:TS_hybrid}
\end{figure}

\begin{figure}
    \centering
     \includegraphics[width = 0.7\textwidth]{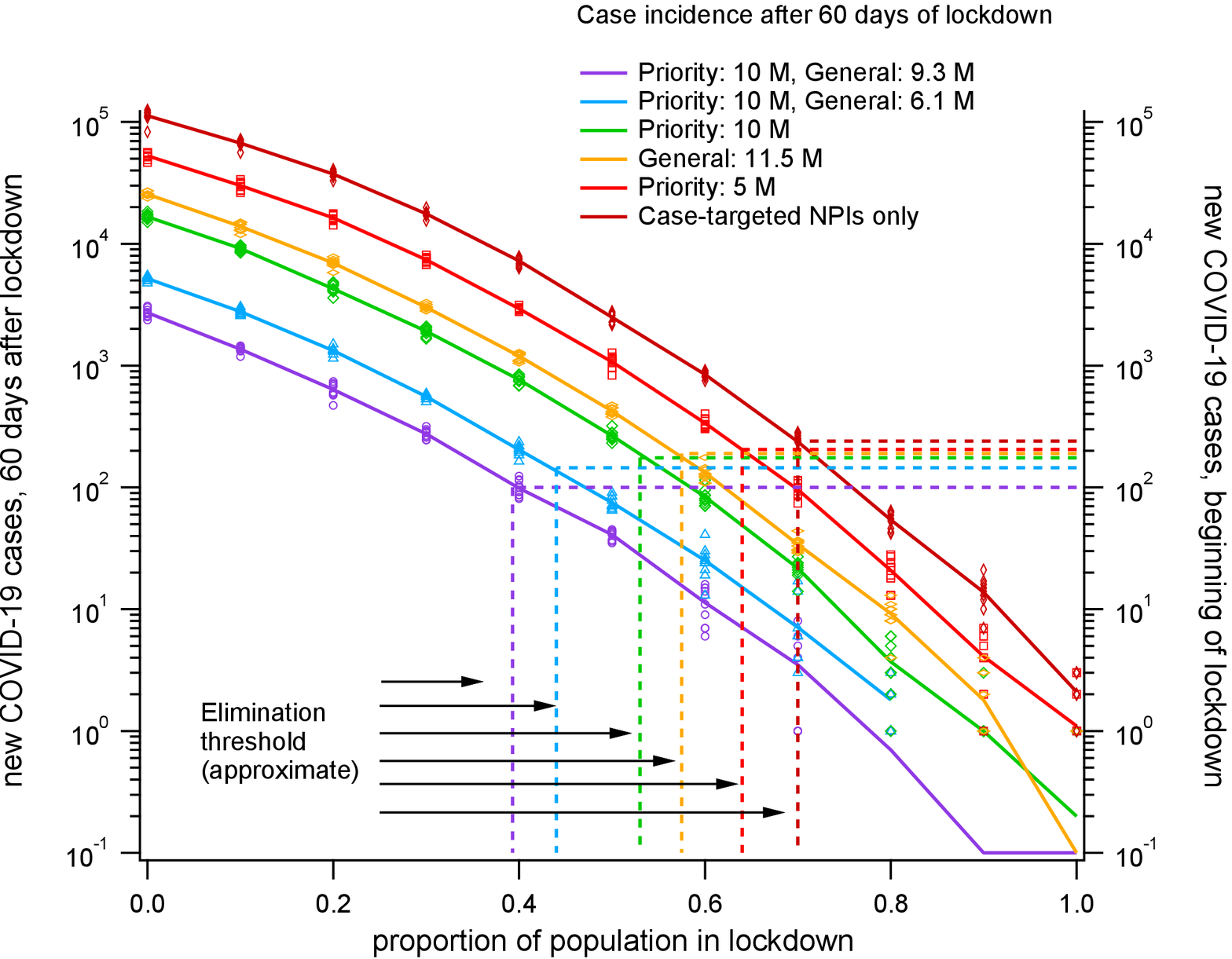}
    \caption{The intensity of lockdown required for gradual elimination of the virus steadily decreases with increasing vaccination levels. Solid lines connect ensemble averages of case incidence 60 days into the lockdown period for each scenario (left y-axis) while the values recorded from each individual simulation are shown as symbols. Each horizontal dashed line corresponds to the average incidence at the onset of lockdown (right y axis) for the vaccination scenario labelled with the same colour. The vertical dashed lines correspond to the approximate proportion of the population in lockdown required for case incidence to decrease, in each vaccination scenario. }
    \label{fig:severity_SD}
\end{figure}

\section*{Discussion}

Stochastic agent-based models have been established as robust tools for tracing fine-grained effects of complex intervention policies in diverse epidemic and pandemic settings~\cite{germann2006mitigation,nsoesie2012sensitivity}. ABM studies have produced policy recommendations developed for the control of COVID-19 outbreaks, which have been adopted broadly by the WHO~\cite{world2020combined}.

Typically, ABMs simulate each individual separately, aiming to account for heterogeneity of demographic and epidemic conditions, as well as details of social interactions and mobility patterns. This approach has a relatively high computational cost, driven by calibration of numerous internal ABM parameters~\cite{ferguson2020impact,chang2020modelling} and reconstruction of mobility patterns. However, ABMs offer an important advantage, combining both behavioural and mechanical adequacy of the model mechanism.  
Behavioural adequacy is verified by comparing simulated and actual epidemic patterns. Mechanical adequacy ensures, in addition, that the natural history of the disease progresses in concordance with the known estimates of incubation periods, serial and generation intervals, and other key parameters. Two ABMs, verified with respect to both behavioural and mechanical adequacy, provided a strong foundation for our study: ACEMod developed to simulate influenza pandemics~\cite{cliff2018nvestigating}, and AMTraC-19 created to simulate COVID-19~\cite{chang2020modelling}.

Our early COVID-19 study~\cite{chang2020modelling}
compared several non-pharmaceutical intervention strategies and identified the minimal  levels of social distancing required to control the pandemic. A compliance rate below 70\% was found to be inadequate for any duration of social distancing, while a compliance at the 90\% level was shown to control the disease within 13-14 weeks, when coupled with other restrictions. In another study we modelled pre-pandemic vaccination and targeted antiviral prophylaxis for influenza pandemics in Australia~\cite{zachreson2020interfering}. However, the COVID-19 pandemic demands new intervention protocols, optimised for given vaccine efficacy and coverage~\cite{bartsch2020vaccine,moore2020modelling}, and accounting for logistical constraints, limited supply and hesitancy.

An early investigation by Bartsch et al. modelled trade-offs between the vaccine efficacy and vaccination coverage, before and during an epidemic in the USA~\citep{bartsch2020vaccine}. The work showed that in order to prevent an epidemic, the efficacy has to be at least 60\% when vaccination coverage is perfect (100\%), and when the latter reduces to 60\%, the necessary efficacy threshold increases to 80\%. During an ongoing epidemic, higher efficacy thresholds were found to be needed to significantly reduce peak severity. 
The study conducted by Moore et al. modelled several vaccine components, in particular distinguishing between the vaccine types that may reduce susceptibility by inhibiting viral transmission (thus, indirectly protecting the individual) and the vaccine types that directly protect only the vaccinated individual by reducing the probability of developing severe symptoms~\citep{moore2020modelling}. Importantly, this study suggested that vaccinating older age groups initially may prevent a second wave within the UK, but only if the vaccine reduces both transmission and disease. Even in this optimistic scenario, only a highly efficacious vaccine, delivered to at least 70\% of the population, was shown to succeed without NPIs. A high level of vaccination was also shown to be required in France, based on the ABM developed by Hoertel et al. who reported that vaccinating only priority groups (e.g., older adults) would be insufficient to lift NPIs~\citep{hoertel2021optimizing}.  Recent evidence from statistical modelling strongly suggests that indirect protection of unvaccinated individuals can indeed occur. However, identification of herd immunity thresholds using statistical models is hindered by many factors including lack of knowledge regarding the levels of natural immunity imparted by recovery from infection, the constantly changing effects of individual behaviour, and the influence of state-imposed mitigation policies \cite{milman2021community}.

Thus, while it is clear that vaccination cannot provide herd immunity without mass population coverage, to what extent coverage requirements can be alleviated by maintaining some (relaxed) level of social distancing and other suppression measures in a highly heterogeneous demographic setting remains an open and country-specific question. This question becomes more complex when several vaccines are considered, each with a different combination of efficacy against susceptibility, infectiousness and disease. 

In attempting to reduce uncertainty of this complex space, we considered different mass vaccination scenarios tailored to the current situation in Australia, while varying the key intervention parameters. As a result, we identified several salient trade-offs between the (pre-pandemic) vaccination and (future) lockdown requirements. Qualitatively, these trade-offs are not dissimilar to those reported by Bartsch et al.~\citep{bartsch2020vaccine}, however, the thresholds which we quantified for Australia are specific for a more refined, hybrid, mass vaccination campaign (with priority and general vaccines). We also reinforce the findings of Moore et al.~\citep{moore2020modelling} with respect to separate components of the vaccine efficacy: future outbreaks become preventable only when the distributed vaccines avert transmission as well as disease. Crucially, no combination of realistic vaccine efficacy values was found to completely eliminate the pandemic threat in Australia without population-scale NPIs, under the feasible vaccination coverage extents that we simulated. This is somewhat different from the results of Moore et al.~\citep{moore2020modelling}, where such an outcome was shown to be theoretically possible in UK, but only by adopting a highly efficacious vaccine. The necessity of partial lockdowns is in concordance with the analysis of Hoertel et al.~\cite{hoertel2021optimizing}, which highlighted the difficulties in lifting NPIs under realistic vaccination strategies considered in France. 

In the main analysis we used a relatively high level of efficacy against infectiousness for the priority vaccine (VEi = 0.684). Recent studies such as that of Harris et al.~\cite{harris2021impact} narrowed the estimates of VEi for BNT162b2  to a lower level (VEi $\approx$ 0.5). In order to verify robustness of our results to changes in VEi between higher and lower levels, we carried out an additional sensitivity analysis which confirmed that the reported outcomes persist across this range of VEi values 
(Fig.~S7).

 There is a remaining uncertainty about the clinical efficacy of ChAdOx1 nCoV-19 (Oxford/AstraZeneca) vaccine  under different vaccine administration schedules. An interim analysis of four randomised controlled trials carried out in three countries (Brazil, South Africa, and the UK) between April and November 2020 suggested a vaccine efficacy of 62.1\% (95\% CI: 41.0--75.7) in participants  who received two standard doses separated by four weeks~\cite{voysey2021safety}. 
When two standard doses were separated by 12 weeks or longer, the vaccine efficacy was observed to be higher at 81.3\% (95\% CI: 60.3--91.2)~\cite{voysey2021single}. However, the follow-up study did not explicitly account for seasonal effects and variation between the trial countries in terms of the epidemic intensity (as well as circulating viral variants). In addition, these trials were not designed to discriminate between vaccine efficacies by dose interval, and therefore, these encouraging ``post-hoc exploratory findings could be biased''~\cite{hung2021single}. Given the reported wide confidence intervals and a possible bias, we 
 considered two settings for  efficacy of the general vaccine. Firstly, 
we adopted a conservative estimate for the efficacy VEc = 0.6 (i.e., 60\%), while varying different components of the efficacy in a broader range. Secondly, we explored a more optimistic setting, VEc = 0.75 (assuming an optimal dose separation regime). We found that our results regarding the lockdown compliance rate required for elimination did not change when we increased the efficacy of the general vaccine from VEc = 0.6 to VEc = 0.75 for the originally planned rollout strategy which followed the allocation ratio of 100:10:1 (for 65+ : 18-64 : <18 age groups, 
Fig.~S8).

The optimistic setting for general vaccine (VEc = 0.75) was also used in sensitivity analysis of 
\begin{enumerate}[(a)]
    \item 
the effects of changing infectiousness efficacy for the priority vaccine from a higher (VEi = 0.684) to a lower level (VEi = 0.5), for the original allocation ratio 100:10:1 
(cf. Figures~S7a~and~S7b), and
   \item the impact of the adjusted rollout strategy based on the revised allocation ratios, while maintaining the more conservative level of efficacy against infectiousness (VEi = 0.5) for the priority vaccine 
	(cf. Figures~S7b~and~S7c).
\end{enumerate}
This comparative analysis confirmed the overall robustness of our results based on the latest available information, including (i) an optimistic setting of efficacy of the general vaccine (VEc = 0.75), (ii) a conservative level of efficacy against infectiousness for the priority vaccine (VEi = 0.5), and (iii) an adopted rollout strategy adjusted for the revised allocation ratios.

Our finding, that herd immunity is not attained even when a large proportion (82\%) of the population is vaccinated, can be explained as a consequence of two correlated sources of heterogeneity. The first is structural, and occurs due to the unavoidable clustering of children in schools and classrooms. The second is imposed by the choice to place children at low priority for immunisation (due to typically low disease severity in this cohort). In Australia, school-aged children (aged from 5 to 18 years), comprise approximately 15\% of the total population (with about 5\% composed of children under the age of 5). Therefore, even with 82\% of the population vaccinated, roughly three quarters of the child population will remain fully susceptible to the virus. Of those who are vaccinated, very few (if any) will receive the priority vaccine. In our ABM, the result is an interconnected subpopulation with high susceptibility. During outbreaks, this produces deviations from the homogeneous approximation in the form of heterogeneous epidemic spread strongly biased towards school-aged children.  While the source of this deviation is produced in our model by the policy decision not to vaccinate large numbers of children, the general result is similar to those related to discretionary differences between clustered social groups with high levels of vaccine scepticism or hesitancy \cite{salathe2008effect}. Therefore, changing health recommendations regarding vaccination of young people may not be sufficient to correct such deviations, which may also be influenced heavily by public opinion and the social clustering of those with similar views.   

It has been widely established that children are at much lower risk of severe COVID-19 disease. However, estimates of transmission rates among children are made difficult by the relatively low rate of symptom expression in young cohorts \cite{flasche2021role}. The current advice from the United States CDC suggests an emerging consensus that transmission rates in young people are similar to those in the adult population \cite{cdc2020pediatric}.  
In our model of COVID-19 transmission, we did not truncate the susceptibility or infectiousness of children but we did assume a lower rate of symptom expression, in line with the available evidence \cite{chang2020modelling}. 

Taken together, these factors mean that age-stratified vaccine prioritisation represents a trade-off: by comprehensively vaccinating older adults with the priority vaccine, the system ensures lower levels of severe disease in the highest-risk cohort, even with low levels of coverage. However, our results demonstrate that this may come at the cost of precluding eventual herd immunity as the virus continues to spread slowly among the (largely asymptomatic) young cohort. Meehan {\it{et al.}} simulated optimised vaccination strategies and demonstrated that in order to achieve herd immunity, those at highest risk of {\it{transmission}} must be targeted for immunisation \cite{meehan2020age}. In our model, infections in children produce a high transmission risk, despite presenting a low risk of symptomatic disease. In this context, and given the current uncertainty with respect to efficacy of COVID-19 vaccines on transmission, our results demonstrate that the Australian strategy trades {\it{possible}} herd immunity for a reduction in severe case load potential.

\subsection*{Limitations and future work}

In comparison with~\cite{chang2020modelling}, our model included a more refined natural history of the disease, calibrated to recent outbreaks which occurred after the first wave in Australia. Nevertheless, the model does not explicitly capture in-hotel quarantine, hospitalisations, and in-hospital transmissions. This limitation is offset by the fact that in Australia's vaccination plan, healthcare and border control professionals are included in the priority vaccination phase, carried out in a pre-pandemic mode. 
We also do not systematically quantify mortality rates, and do not elaborate on the expanding disease surveillance capacity and standard clinical pathways in Australia~\cite{moss2020modelling}. 

As we pointed out, no herd immunity is attainable under currently feasible conditions. This outcome has several ramifications beyond the direct consequences of future partial lockdowns. On the one hand, the main reason for the inadequate collective immunity is the existence of highly clustered, networked communities (e.g., educational, religious and community groups, etc.). This highlights the need for a more sophisticated simulation of contact networks, in addition to the workplace/school environments generated from the census data. On the other hand, the lack of herd immunity may affect the behaviour of the ``free-riders'' who typically exploit the collective protection of mass immunisation while not committing to the vaccination themselves. This may create a feedback loop, reducing vaccine hesitancy in the near- to mid-term, and generating long-term oscillatory dynamics in vaccine adoption~\cite{chang2020impact}. 

Another caveat is that our ABM population is matched to the latest Australian Census data, which was collected in 2016. This produces a model population of 23.4M individuals, which is smaller than the current Australian population by approximately 2M people. Due to this discrepancy, the coverage proportions defined for fixed numbers of vaccines are slightly inflated in our simulations (by about 8\%), relative to what would be achieved with the current population count. Because we did not identify a threshold in epidemic severity as a function of vaccine coverage in our hybrid scenarios, we do not expect this discrepancy to qualitatively alter our results. 

Finally, our simulations treat the entire population as initially susceptible, not accounting for the influence of pre-existing immunity in the population produced by previous waves of COVID-19 in Australia. The effective suppression of these previous outbreaks has kept cumulative confirmed case totals below 1\% of the total population, so we do not expect this simplifying assumption to influence our results.

\subsection*{Conclusion}

In this work, we extended a high-resolution agent-based model of COVID-19 mitigation and control to simulate the effects of coupled vaccination and non-pharmaceutical strategies on future outbreaks in Australia. We found that, combined with case-targeted interventions, a completed mass-vaccination campaign using both a 90\% effective vaccine for priority populations and a 60\% effective vaccine for the general population would dramatically slow viral spread but would not produce herd immunity. If a community transmission outbreak were to occur during or after the vaccination campaign, population-scale non-pharmaceutical interventions (i.e., lockdown) would be necessary to curb transmission. In our simulations, the required extent and duration of these measures decreased gradually with the level of vaccination coverage obtained by the time of the outbreak. For realistic endpoint conditions, with 82\% of the population vaccinated, the required lockdown intensity decreased by 43\% and and initial epidemic growth rate decreased by 52\%. Due to coupling between these two factors (broadening of the epidemic curve and increased effectiveness of nonpharmaceutical interventions), the severity of epidemics as measured by the peak number of new cases over a 24hr period decreased by up to two orders of magnitude under plausible mass-vaccination campaign endpoint conditions.

With respect to the prognosis for future outbreaks of COVID-19 in Australia, several important questions remain unaddressed by our study. Because we did not model the effect of vaccination on hospital case load, medical infrastructure, and mortality, our results do not directly inform estimates of the trade-off between the socioeconomic costs of lockdown and the human cost of allowing a low level of COVID-19 transmission. Finally, at the time of writing, several SARS-CoV-2 variants of concern have been identified. Among them, variant B.1.1.7 (Alpha), which emerged in the United Kingdom and has spread globally, was associated with increased transmission  
potential: specifically, its $R_0$ was estimated to be 43-90\% (95\% CrI: 38--130\%) higher than reproduction number of preexisting variants~\cite{davies2021estimated}. Variant B.1.351 (Beta), originally identified in South Africa, has been provisionally associated with lower rates of neutralisation by polyclonal antibodies, and variant P.1 (Gamma), thought to have originated in Brazil, has been associated with a major outbreak in a population thought to be effectively immune \cite{greaney2021comprehensive,sabino2021resurgence}. Most recently, variant B.1.617.2 (Delta), first detected in India, has become the dominant strain in many countries, increasing risk of household transmission by an approximately 60\% compared to the Alpha variant~\cite{phe-11-june-2021}. Given the currently low level of understanding about the implications of these variants in future outbreak scenarios, our results should be viewed as optimistic guidelines that assume the continuing vaccination efforts can keep pace with the evolution of SARS-CoV-2.

\section*{Contributors}
CZ and MP conceived and co-supervised the study, designed the computational experiments and drafted the original Article. CZ and OMC developed the agent-based model. CZ implemented intervention strategies. All authors developed and calibrated the COVID-19 epidemiological model, and performed sensitivity analysis. SLC and CZ   carried out the computational experiments and analysis, verified the underlying data, and prepared all figures. All authors had full access to all the data in the study.  All authors contributed to the editing of the Article, and read and approved the final Article.

\section*{Declaration of interests}
We declare no competing interests.

\section*{Data sharing}

Post-processing source data and supplementary data are provided with this Article. The full data can be made available to approved bona fide researchers after their host institution has signed a Data Access/Confidentiality Agreement with the University of Sydney. Mediated access will enable data to be shared and results to be confirmed without unduly compromising the University’s ability to commercialise the software. To the extent that this data sharing does not violate the commercialisation and licensing agreements entered into by the University of Sydney, the data will be made publicly available after the appropriate licensing terms agreed. 

\section*{Acknowledgments} 
This work was partially supported by the Australian Research Council grant DP200103005 (MP and SLC). Additionally, CZ is supported in part by National Health and Medical Research Council project grant (APP1165876). 
AMTraC-19 is registered under The University of Sydney’s invention disclosure CDIP Ref. 2020-018. We are thankful for support provided by High-Performance Computing (HPC) service (Artemis) at the University of Sydney.

\FloatBarrier

\bibliography{references.bib}

\newpage

\newcommand{\beginsupplement}{%

 \setcounter{table}{0}
   \renewcommand{\thetable}{S\arabic{table}}%
   
     \setcounter{figure}{0}
      \renewcommand{\thefigure}{S\arabic{figure}}%
      
      \setcounter{page}{1}
      \renewcommand{\thepage}{S\arabic{page}} 
      
      \setcounter{section}{0}
      \renewcommand{\thesection}{S\arabic{section}}
      
      \setcounter{equation}{0}
      \renewcommand{\theequation}{S\arabic{equation}}
     }


\beginsupplement


%



\begin{centering}
\section*{\Large{Supplementary Material for: \\ 
 \ \\ How will mass-vaccination change COVID-19 lockdown requirements in Australia?\\
\vspace{0.5cm}
\large{Cameron~Zachreson$^{1,2}$, Sheryl~L.~Chang$^{1}$, Oliver~M.~Cliff$^{1,3}$,  Mikhail~Prokopenko$^{1,4}$}\\
\vspace{0.5cm}
\small{$^{1}$  Centre for Complex Systems, Faculty of Engineering, The University of Sydney, Sydney, NSW 2006, Australia\\
$^{2}$  School of Computing and Information Systems, The University of Melbourne, Parkville, VIC 3052, Australia\\
$^{3}$  Centre for Complex Systems, School of Physics, Faculty of Science, The University of Sydney, Sydney, NSW 2006, Australia\\
$^{4}$  Sydney Institute for Infectious Diseases, The University of Sydney, Westmead, NSW 2145, Australia\\ }}}
\end{centering}
\vspace{1cm}

\subsection*{Constraints on vaccine efficacy}

The efficacy of a vaccine can be decomposed into three components:
\begin{itemize}
    \item{the efficacy for susceptibility ($\VEs$) which determines the level of immunity induced in susceptible individuals}
    \item{the efficacy for disease ($\VEd$) which determines protection against symptomatic illness if a vaccinated individual is infected}
    \item{efficacy against infectiousness ($\VEi$) which determines how contagious a vaccinated individual will be if they become infected}
\end{itemize}
The clinical efficacy ($\VEc$), which is measured in stage-3 clinical trials and corresponds to the reported efficacy numbers for both vaccines, is a function of $\VEs$ and $\VEd$, neither of which are directly measured:
\begin{equation}\label{Eq_VEc_supp}
    \VEc = \VEd + \VEs  - \VEs\VEd,
\end{equation}
while the efficacy for transmission, $\VE$, which is the value of interest for computing herd immunity is a function of vaccine coverage ($c$), $\VEi$, $\VEs$, and $\VEd$. Because $\VEi$ is not constrained by $\VEc$ through Equation~\ref{Eq_VEc_supp}, the population-level effectiveness is currently uncertain for both vaccines. 

In the worst-case scenario, the factor $\VEd$ would account for 100\% of the clinical efficacy ($\VEd = \VEc$, $\VEs = 0$), and the efficacy against transmission would be negligible ($\VEi = 0$), which would leave all unvaccinated individuals unprotected. In this scenario, herd immunity is not possible per se (though complete coverage would provide clinical protection to the entire population). In the best-case scenario, efficacy against symptoms would be maximised ($\VEd = \VEc$), protecting all vaccinated individuals regardless of their infection status, while the unknown parameter $\VEi$ would take a value of 1, and all transmission from vaccinated individuals would cease, protecting both vaccinated and unvaccinated subpopulations. In reality, the true efficacy values will lie between these extremes, and may be correlated. For example, because symptoms may increase the contagiousness of those infected, $\VEd$ can positively influence $\VEi$. On the other hand, symptoms can also lead to case detection and behavioural change, which may produce a negative relationship between $\VEd$ and $\VEi$. 

\subsection*{Vaccine efficacy: homogeneous approximation}

Here, we give a homogeneous approximation for vaccine efficacy and the associated herd immunity thresholds as functions of $\VEs$, $\VEi$, and $\VEd$. We derive an expression for overall reduction to force of infection provided by a vaccination program with coverage $c \in [0, 1]$ as follows:
\begin{equation}
    \VE = 1 - F_{vax} / F_o \,, 
\end{equation}
where $F_o$ is proportional to the overall force of infection when none of the population is vaccinated, and $F_{vax}$ is the relative reduction to force of infection produced through vaccination. For a given prevalence of infection $p(\text{infected}) = N_{\text{infected}}/N_{\text{tot}} << 1 = 1 - p(\text{susceptible})$, and assuming a negligible recovered population, we compute $F_o$ as: 
\begin{equation}\label{Eq_Fo}
    F_o = \beta_o^2~p(\text{infected})~p(\text{susceptible})\,,
\end{equation}
in which $\beta_o$ is the average force of infection produced by an infected individual who is unvaccinated:
\begin{equation}
    \beta_o = \beta (p_s + a(1-p_s))\,
\end{equation}
in which $\beta$ is the force of infection from a symptomatic individual, $p_s$ is the probability of expressing symptoms if infected and unvaccinated (in this work we assumed a symptomatic fraction of 2/3) and $a$ is the factor by which force of infection is reduced if an infected agent is asymptomatic (for this work $a = 0.3$). The $p(\text{infected}) p(\text{susceptible})$ term in Eq.~\eqref{Eq_Fo} captures the potential for interaction between susceptible and infected individuals, the potential for transmission given interaction is given by one of the $\beta_o$ terms, and the infectious potential of the newly infected individual, should transmission take place, is given by the second $\beta_o$ term. 

The relative force of infection with vaccination, $F_{\text{vax}}$, can be found similarly by summing the potential for contact and transmission between individuals with different vaccination status: 
\begin{equation}\label{F_vax}
    F_{\text{vax}} = F_{\text{unvax} \rightarrow \text{unvax}} + F_{\text{unvax} \rightarrow \text{vax}} + F_{\text{vax} \rightarrow \text{unvax}} + F_{\text{vax} \rightarrow \text{vax.}}\,,
\end{equation}
where
\begin{equation}
    F_{\text{unvax} \rightarrow \text{vax}} = \beta \beta_{\text{vax}} (1 - \VEs) ~p(\text{infected,}~\text{unvaccinated}) ~p(\text{susceptible,}~\text{vaccinated})\,,
\end{equation}
in which
\begin{equation}\label{Eq_betavax}
    \beta_{\text{vax}} = \beta(1 - \text{$\VEi$})~ [ p_s (1-\VEd) + a (1 - p_s (1-\VEd))]\,,
\end{equation}
is the average force of infection from a vaccinated individual who becomes infected. Note that both $\VEi$ and $\VEd$ play a role in computing $\beta_{\text{vax}}$. The remaining terms of Eq.~\eqref{F_vax} are: 
\begin{equation}
    F_{\text{vax} \rightarrow \text{unvax}} = \beta_{\text{vax}}~\beta_o ~p(\text{infected,}~\text{vaccinated})~p(\text{susceptible,}~\text{unvaccinated})\,,
\end{equation}
and
\begin{equation}
    F_{\text{vax} \rightarrow \text{vax}} = \beta_{\text{vax}}^2~(1-\VEs) ~p(\text{infected,}~\text{vaccinated})~p(\text{susceptible,}~\text{vaccinated})\,.
\end{equation}

The joint probabilities
\begin{itemize}
    \item $p(\text{infected,}~\text{unvaccinated})$,
    \item $p(\text{infected,}~\text{vaccinated})$,
    \item $p(\text{susceptible,}~\text{unvaccinated})$, and
    \item $p(\text{susceptible,}~\text{vaccinated})$
\end{itemize}
can be represented as products of the conditional and independent probabilities of infection, vaccination, and infection given vaccination: 
\begin{equation}
    p(\text{infected,}~\text{unvaccinated}) = p(\text{infected}~|~\text{unvaccinated}) ~ p(\text{unvaccinated})\,,
\end{equation}
\begin{equation}
    p(\text{susceptible,}~\text{unvaccinated}) = p(\text{susceptible}~|~\text{unvaccinated}) ~ p(\text{unvaccinated})\,,
\end{equation}
\begin{equation}
    p(\text{infected,}~\text{vaccinated}) = p(\text{infected}~|~\text{vaccinated}) ~ p(\text{vaccinated})\,,
\end{equation}
\begin{equation}
    p(\text{susceptible,}~\text{vaccinated}) = p(\text{susceptible}~|~\text{vaccinated}) ~ p(\text{vaccinated})\,,
\end{equation}
where 
\begin{equation}
    p(\text{vaccinated}) = c\,,
\end{equation}
\begin{equation}
    p(\text{unvaccinated}) = 1 - c\,,
\end{equation}
\begin{equation}
    p(\text{susceptible}~|~\text{unvaccinated}) = 1 - p(\text{infected}~|~\text{unvaccinated})\,,
\end{equation}
\begin{equation}
    p(\text{susceptible}~|~\text{vaccinated}) = 1 - p(\text{infected}~|~\text{vaccinated})\,,
\end{equation}

\begin{equation}
    p(\text{infected}~|~\text{unvaccinated}) = \frac{p(\text{infected}) - p(\text{infected}~|~\text{vaccinated}) ~ p(\text{vaccinated})}{1 - p(\text{vaccinated})}\,,
\end{equation}

and
\begin{equation}
p(\text{infected}~|~\text{vaccinated}) = p(\text{infected})~(1-\VEs)~\frac{\beta_{\text{vax}}}{\beta}\,,
\end{equation}
where the ratio $\beta_{\text{vax}}/\beta$ gives the effect of $\VEd$ and $\VEi$ on infectiousness of vaccinated individuals (Eq.~\eqref{Eq_betavax}). Here, the state denoted ``infected'' would more precisely be described as ``infectious'', because $\VEi$ and $\VEd$ reduce transmission potential without reducing the infection probability of vaccinated individuals. 

Assuming that the effective reproductive ratio $R$ is proportional to the force of infection $F_{vax}$, it can be computed from the total efficacy, $\VE$, as
\begin{equation}\label{R_eff_supp}
    R = R_0(1 - \VE),
\end{equation}
where $R_0$ is the basic reproductive number for the epidemic, which we set to $R_0 = 2.75$ in the present work. Using $R_0 = 2.75$, Fig.~\ref{fig:coverage_threshold} demonstrates the coverage threshold for herd immunity as a function of $\VEi$, $\VEs$, and $\VEd$ given $\VEc = 0.6$ (Fig.~\ref{fig:coverage_threshold}a), and $\VEc = 0.9$ (Fig.\ref{fig:coverage_threshold}b).

The homogeneous approximation corresponds qualitatively to the results of the ABM which demonstrates a dramatic decrease in growth rate after increasing coverage crosses the $R = 1$ boundary estimated by evaluating Eq.~\eqref{R_eff_supp} for $R_0 = 2.75$ (Fig.~\ref{fig:coverage_threshold_ABM_gr}, Fig.~\ref{fig:coverage_threshold_ABM_pp}). However, due to heterogeneity in population structure and vaccine allocation we still observed substantial epidemic growth for parameter combinations which correspond to $R < 1$ based on the homogeneous approximation. For the priority vaccine ($\VEc = 0.9$), the threshold computed through the homogeneous approximation matched the observed nonlinear decrease in growth rate more closely (Fig.~\ref{fig:coverage_threshold_ABM_gr}b), but finite epidemics were still observed for coverage $\leq 79\%$. The boundary computed by the homogeneous approximation should therefore be viewed as an optimistic estimate of coverage threshold. 

Epidemic severity as measured by peak prevalence follows the expected trend based on the homogeneous approximation (Fig. \ref{fig:coverage_threshold_ABM_pp} and Fig. \ref{fig:coverage_threshold_ABM_CTNPI_pp}). However, for simulations with growth rates less than $\approx 0.06$, peak prevalence was not reached within the simulation time frame of 194 days (these scenarios are indicated as open black circles in Fig. \ref{fig:coverage_threshold_ABM_pp} and Fig. \ref{fig:coverage_threshold_ABM_CTNPI_pp}). The correspondence between the herd immunity threshold computed through the homogeneous approximation and the tendency for simulations to reach peak prevalence within 194 days is an artefact of the choice of simulation horizon. 

\begin{figure}
    \centering
    \includegraphics[width=0.8\textwidth]{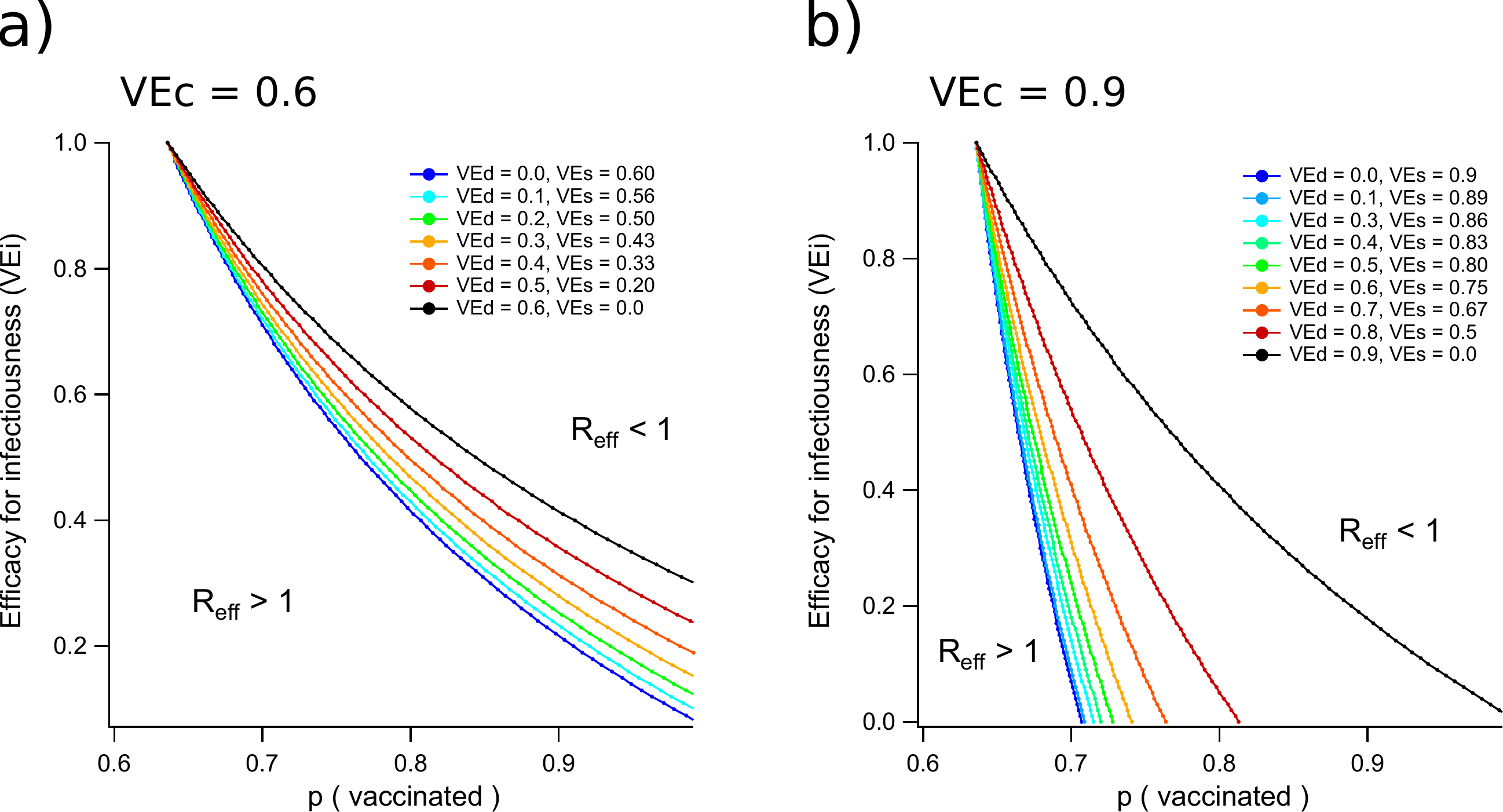}
    \caption{Herd immunity thresholds as functions of efficacy for infectiousness ($\VEi$) and vaccine coverage ($p(\text{vaccinated})$) computed through a homogeneous approximation of vaccine effectiveness taking into account three forms of vaccine efficacy (\VEs, \VEd, and \VEi). Subplot (a) shows thresholds for the general vaccine, with clinical efficacy $\VEc = 0.6$, while subplot (a) shows thresholds for the priority vaccine with clinical efficacy $\VEc = 0.9$.}
    \label{fig:coverage_threshold}
\end{figure}

\begin{figure}
    \centering
    \includegraphics[width=\textwidth]{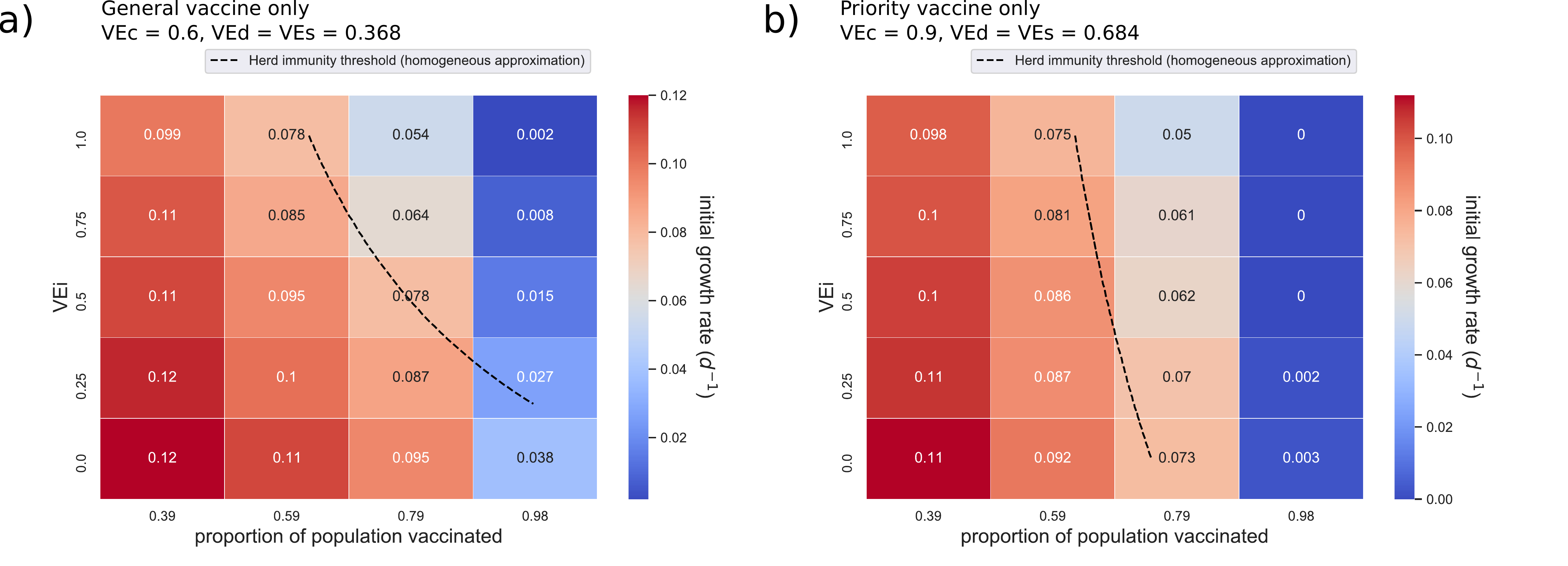}
    \caption{Alignment of ABM results with homogeneous approximation of herd immunity thresholds for the general (a) and priority (b) vaccines. The growth rates were computed from results generated by the ABM over a range of values for coverage and vaccine efficacy against infectiousness. The dashed black lines give the coverage thresholds for herd immunity estimated by the homogeneous approximation. Here, central values of efficacy against disease and susceptibility were used ($\VEd = \VEs = 1 - \sqrt{1 - \VEc}$).}
    \label{fig:coverage_threshold_ABM_gr}
\end{figure}

\begin{figure}
    \centering
    \includegraphics[width=\textwidth]{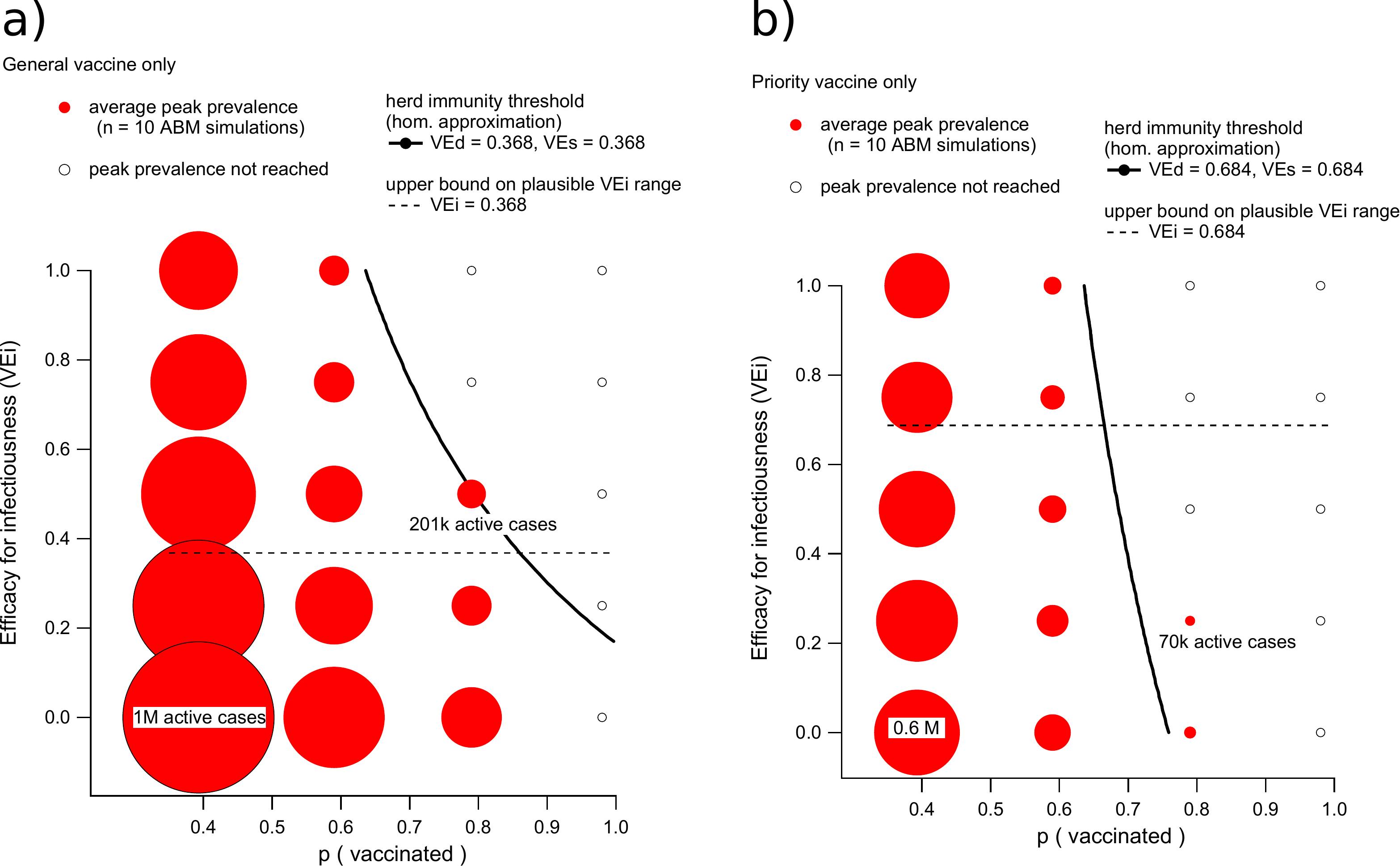}
    \caption{Alignment of ABM results with homogeneous approximation of herd immunity thresholds for the general (a) and priority (b) vaccines. The peak prevalence values shown here (red circles) generated by the ABM over a range of values for coverage and vaccine efficacy against infectiousness. The solid black lines give the coverage thresholds for herd immunity estimated by the homogeneous approximation, and the dashed lines illustrate conservative practical upper bounds on $\VEi$. Here, central values of efficacy against disease and susceptibility were used ($\VEd = \VEs = 1 - \sqrt{1 - \VEc}$). Open black dots indicate simulation sets that did not consistently reach a defined prevalence maximum before the endpoint of 194 days.}
    \label{fig:coverage_threshold_ABM_pp}
\end{figure}

\begin{figure}
    \centering
    \includegraphics[width=\textwidth]{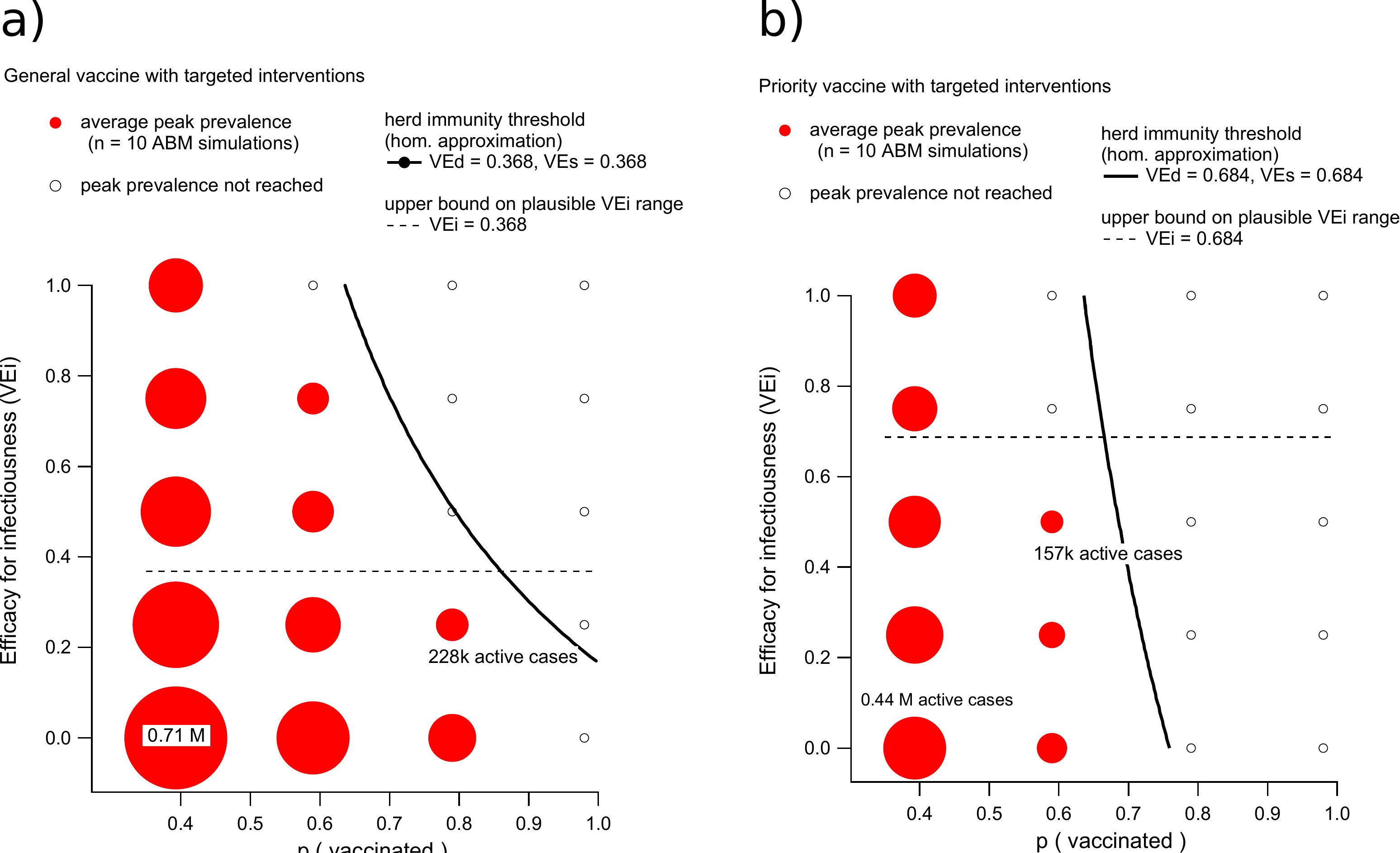}
    \caption{Alignment of ABM results with homogeneous approximation of herd immunity thresholds for the general (a) and priority (b) vaccines, combined with case-targeted nonpharmaceutical interventions. The peak prevalence values shown here (red circles) generated by the ABM over a range of values for coverage and vaccine efficacy against infectiousness. The solid black lines give the coverage thresholds for herd immunity estimated by the homogeneous approximation, and the dashed lines illustrate conservative practical upper bounds on $\VEi$. Here, central values of efficacy against disease and susceptibility were used ($\VEd = \VEs = 1 - \sqrt{1 - \VEc}$). Open black dots indicate simulation sets that did not consistently reach a defined prevalence maximum before the endpoint of 194 days.}
    \label{fig:coverage_threshold_ABM_CTNPI_pp}
\end{figure}

\FloatBarrier

\subsection*{Model calibration}

Primary calibration was performed by first tuning disease parameters $\kappa$, and the bounds of $T_{\text{symp}}$ for a basic reproductive number falling within the range specified in literature reports \cite{billah2020reproductive}. To compute $R_0$ for a given set of parameters, we performed an age-group biased micro-simulation Monte Carlo estimate of secondary cases produced by a typical index case as described in our previous studies \cite{zachreson2020interfering,chang2020modelling}. 

We then tuned the parameters defining case-targeted NPIs to match the early incidence data issued in Australia during the initial wave of COVID-19 in March 2020. The resulting incidence growth rates generated by our ABM (growth rate $\approx$ 0.118) lie between the plausible values estimated for the first wave of COVID-19 (growth rate $\approx$ [0.1, 0.2] Fig.~\ref{fig:calibration_incidence}a). If the first three cases are ignored when determining the growth rate for the first wave, the calculated rate doubles (Fig.~\ref{fig:calibration_incidence}a), but provides a better fit to the remaining case data. It is unclear whether this discrepancy occurred due to ineffective case ascertainment (due to e.g., delays to initial surveillance efforts), or to stochastic die-out of the outbreak that produced the first recorded cases. 

On the other hand, the growth rate produced by our ABM closely matches the rate estimated for the second wave, which began in early June, 2020, was confined to the state of Victoria, and occurred mostly within the urban area of Greater Melbourne \cite{zachreson2021risk} (growth rate $\approx$ 0.123, Fig.~\ref{fig:calibration_incidence}b). While our ABM simulates the entirety of Australia, early cases mostly arise in urban centres which contain the international airports from which importations are generated \cite{cliff2018nvestigating}. Therefore, a close match between the growth rates generated by our model and the rate observed during the beginning of the second wave in Greater Melbourne indicates that our simulations produce reasonable approximations to the early disease dynamics of an Australian outbreak.

The parameters defining population-scale NPIs (lockdown) were chosen to match the peak incidence and prevalence data produced during the first wave, which was suppressed with a national-scale lockdown (Fig.~\ref{fig:IP_calibration}). While the initial incidence growth dynamics of our model are qualitatively similar to observations from the 1st wave (Fig.~\ref{fig:calibration_incidence}b, Fig.~\ref{fig:IP_calibration}a), and well-matched to those observed during the second wave (Fig.~\ref{fig:calibration_incidence}b), there are some aspects of the observed data from the first wave that are not reproduced well by our model. In particular, simulated case prevalence is substantially lower than the number of active cases reported during the first wave. This is a direct consequence of our decision to model the infectious period after symptom onset in accordance to the period of replication competent viral shedding rather than the period over which a case may test positive (which can be longer by 1-2 weeks \cite{cdc2021interim}). Therefore, in our model a case recovers and is no longer included in the prevalence count substantially earlier than it would be removed from an active case count ascertained through PCR tests. In addition, the cumulative incidence produced by our model is higher than what was observed in the first wave during March 2020 (Fig.~\ref{fig:IP_calibration}b). However, this discrepancy only becomes substantial during the period after lockdown was implemented, so it can be interpreted as the consequence of our conservative estimates in the efficacy of lockdown on transmission in various contexts (Table \ref{tab:control_params_lockdown}). We chose conservative values for these parameters, so that our estimates of lockdown thresholds would err on the side of caution.

\setlength{\arrayrulewidth}{0.5mm}
\setlength{\tabcolsep}{18pt}
\renewcommand{\arraystretch}{1.5}
\begin{table}
    \centering
    \resizebox{\textwidth}{!}{
    \begin{tabular}{c|c|c|c}
        parameter & value & distribution & notes \\
        \hline \hline
         $\kappa$ & 2.4 & NA & global transmission scalar \\\hline
         $T_{\text{inc}}$ & 5.5 days (mean)  & lognormal($\mu = 1.62,~\sigma = 0.418$)  & incubation period \\\hline
         $T_{\text{symp}}$ & 10.5 days (mean)  & uniform [7, 14]  & symptomatic (or asymptomatic) period \\\hline
         $a$ & 0.3  & NA  & asymptomatic transmission scalar \\\hline
         $p_{\text{ symptomatic}}~|~\text{adult}$ & 0.67  & NA & probability of symptoms (age $<$ 18)  \\ \hline
         $p_{\text{ symptomatic}}~|~\text{child}$ & 0.134  & NA & probability of symptoms (age 18+)  \\ \hline
         $p_{\text{ detect}}~|~\text{symptomatic}$ & 0.227  & NA & daily case detection prob. (symptomatic)  \\ \hline
         $p_{\text{ detect}}~|~\text{asymptomatic}$ & 0.01  & NA & daily case detection prob. (asymptomatic)  \\ \hline
    \end{tabular}}
    \caption{Key control parameters for COVID-19 transmission model. }
    \label{tab:control_params_transmission}
\end{table}

\setlength{\arrayrulewidth}{0.5mm}
\setlength{\tabcolsep}{18pt}
\renewcommand{\arraystretch}{1.5}
\begin{table}
    \centering
    \resizebox{\textwidth}{!}{
    \begin{tabular}{c|c|c|c}
        parameter & value & distribution & notes \\
        \hline \hline
         $p_{\text{CI}}$ & 0.7  & NA  & case isolation compliance rate \\\hline
         $p_{\text{HQ}}$ & 0.5 & NA & home quarantine compliance rate \\\hline
         $T_{\text{HQ}}$ & 14 d & NA & home quarantine duration \\ \hline
         $f_{\text{home}}(\text{HQ})$ & 2  & NA  & NPI transmission scalar (HQ, home) \\\hline
         $f_{\text{community}}(\text{HQ})$ & 0.25  & NA  & NPI transmission scalar (HQ, community) \\\hline
         $f_{\text{workplace}}(\text{HQ})$ & 0.25  & NA  & NPI transmission scalar (HQ, workplace) \\\hline
         $f_{\text{home}}(\text{CI})$ & 1  & NA  & NPI transmission scalar (CI, home) \\\hline
         $f_{\text{community}}(\text{CI})$ & 0.25  & NA  & NPI transmission scalar (CI, community) \\\hline
         $f_{\text{workplace}}(\text{CI})$ & 0.25  & NA  & NPI transmission scalar (CI, workplace) \\\hline
    \end{tabular}}
    \caption{Key control parameters for targeted NPIs. NPI transmission scalars multiply the force of infection produced by infected individuals in the specified contexts. Compliance rates determine the proportion of individuals who act in accordance with the specified measures (case isolation, CI; home-quarantine of household contacts, HQ).}
    \label{tab:control_params_NPIs}
\end{table}

\setlength{\arrayrulewidth}{0.5mm}
\setlength{\tabcolsep}{18pt}
\renewcommand{\arraystretch}{1.5}
\begin{table}
    \centering
    \resizebox{\textwidth}{!}{
    \begin{tabular}{c|c|c|c}
        parameter & value & distribution & notes \\
        \hline \hline
         $p_{\text{LD}}$ & variable  & [0, 1]  & lockdown compliance rate \\\hline
         $T_{\text{LD}}$ & 91 d & NA & lockdown duration \\ \hline
         $f_{\text{home}}(\text{LD})$ & 1  & NA  & NPI transmission scalar factor (LD, home) \\\hline
         $f_{\text{community}}(\text{LD})$ & 0.25  & NA  & NPI transmission scalar factor (LD, community) \\\hline
         $f_{\text{workplace}}(\text{LD})$ & 0.1  & NA  & NPI transmission scalar (LD, workplace) \\\hline
         $\text{AR}_{\text{trigger}}(\text{LD})$ & 2000  & NA  & cum. incidence triggering population-level NPIs \\\hline
    \end{tabular}}
    \caption{Key control parameters for population-level NPIs (lockdown, LD).}
    \label{tab:control_params_lockdown}
\end{table}

\setlength{\arrayrulewidth}{0.5mm}
\setlength{\tabcolsep}{18pt}
\renewcommand{\arraystretch}{1.5}
\begin{table}
    \centering
    \resizebox{\textwidth}{!}{
    \begin{tabular}{c|c|c|c}
        parameter & value from ABM & target value & notes \\
        \hline \hline
         $R_0$ & 2.75 [2.71, 2.80] & 2.9 [2.39, 3.44] & basic reproductive ratio~\cite{billah2020reproductive} \\\hline
         $T_{\text{gen}}$ & 7.14 [7.05, 7.23] & 7.0 days [5.8, 8.1]  & generation/serial interval~\cite{wu2020estimating} \\\hline
         growth rate & 0.118 [0.110, 0.127] & 0.10 [0.097, 0.103]  & growth rate of case incidence (1st wave) \\\hline
         growth rate & 0.118 [0.110, 0.127] & 0.201 [0.170, 0.233]  & growth rate of case incidence (1st wave, from day 21) \\\hline
         growth rate & 0.118 [0.110, 0.127] & 0.123 [0.102, 0.143]  & growth rate of case incidence (2nd wave, VIC) \\\hline
         peak prevalence & 2790, range [2623, 3059] & 4935  & peak case prevalence (1st wave) \\\hline
         peak incidence & 377, range [336, 415] & 497  & peak case incidence (1st wave) \\\hline
    \end{tabular}}
    \caption{Calibration targets for key model outputs. }
    \label{tab:calibration_targets}
\end{table}

\begin{figure}
    \centering
    \includegraphics[width=\textwidth]{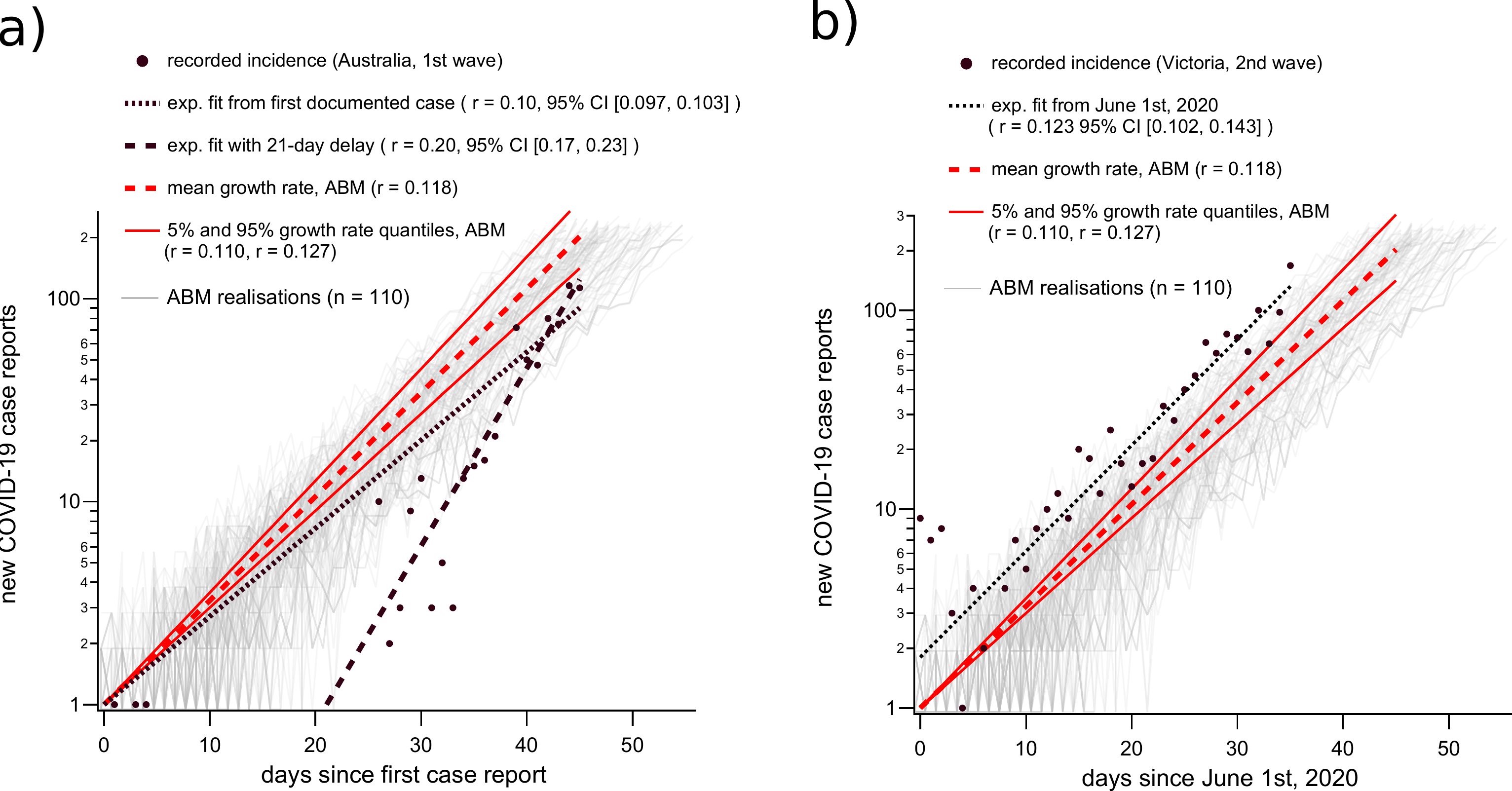}
    \caption{The growth rate computed by the ABM with case-targeted NPIs approximately matches the rate computed for the 2nd wave of COVID-19 in Victoria. Incidence growth rates computed by the model are compared to case data recorded by the Australian Department of Health for the 1st wave of COVID-19 (beginning on Feb. 3rd, 2020) (a), and case data recorded by the Victorian Department of Health and Human Services for the 2nd wave of COVID-19 (b).}
    \label{fig:calibration_incidence}
\end{figure}

\begin{figure}
    \centering
    \includegraphics[width=\textwidth]{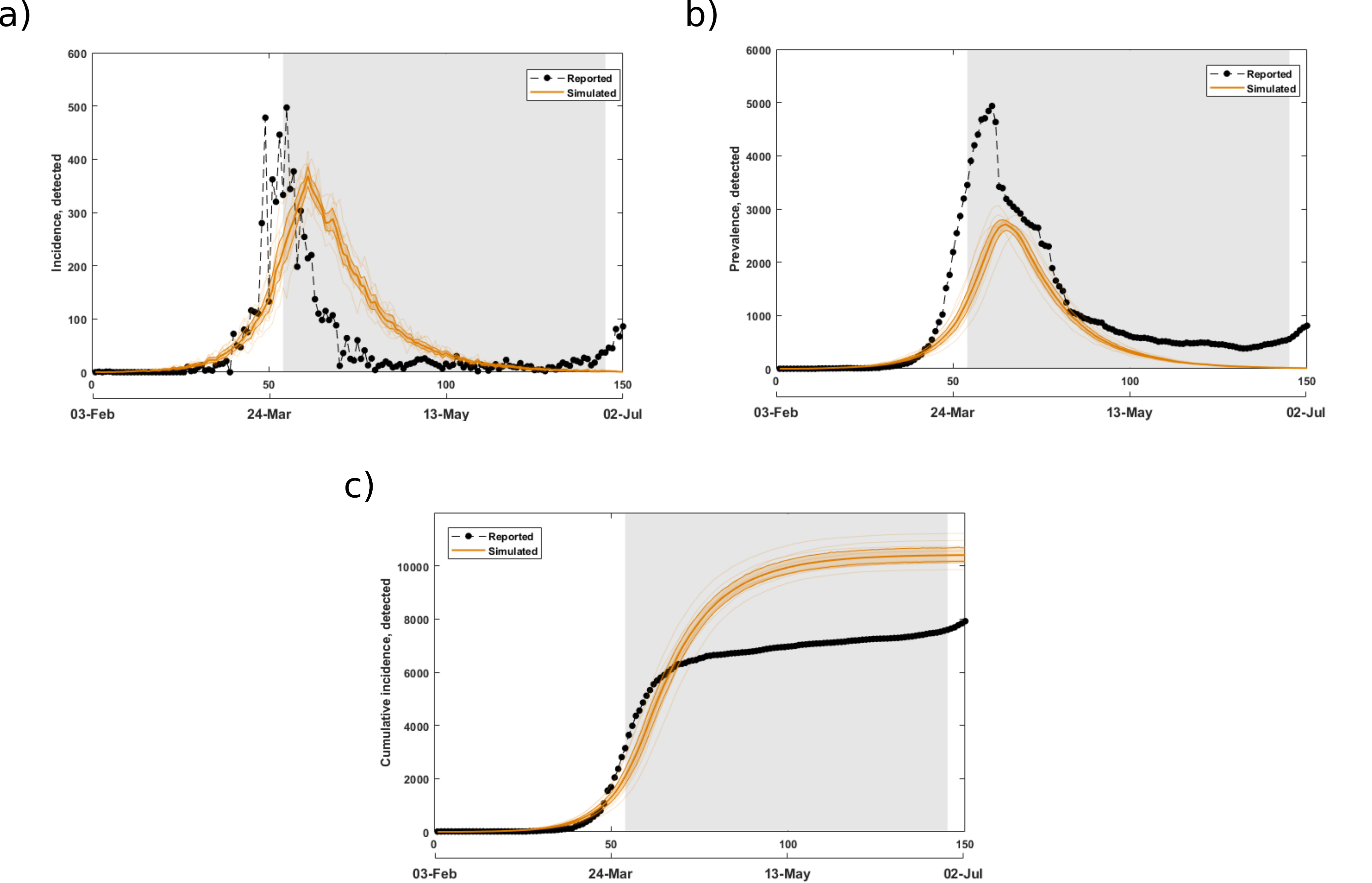}
    \caption{Alignment of model results with recorded case data from the first wave of COVID-19 in Australia. Daily case incidence is shown in (a), with reported case data shown as black dots connected by dashed lines and mean case incidence produced by the ABM shown as a solid yellow line (shaded bands represent 95\% bootstrap CI bounds). The output of individual instances of the ABM (n = 10) are shown as transparent yellow traces. Case prevalence (active case counts) are shown in (b), and cumulative case incidence is shown in (c). The grey shaded region corresponds to the lockdown period used in the ABM, which utilised a lockdown compliance level  of 90\% to match the conditions used in our previous work \cite{chang2020modelling}.}
    \label{fig:IP_calibration}
\end{figure}
\FloatBarrier

\subsection*{Sensitivity of lockdown threshold to vaccine efficacy and age-specific priority}
Figures \ref{fig:SX_sensitivity_trajectories} and \ref{fig:SX_sensitivity_SD_thresh} present results of a sensitivity analysis with respect to vaccine efficacy and vaccine allocation priority among the three age groups used in our model. Vaccine efficacies and vaccine age-group priority specifications are given in the figure legends. Rollout ratios listed in the legends $[x_1:x_2:x_3]$ correspond to age groups [65+ : 18-64 : <18], the revised rollout numbers [2547:30000:1000] were chosen in order to invert the priority vaccine distribution between 65+ and 18-64 age groups, while holding constant the overall number of vaccines distributed to each age group as allocated by the original 100:10:1 priority system.

\begin{figure}[h]
    \centering
    \includegraphics[width=0.8\textwidth]{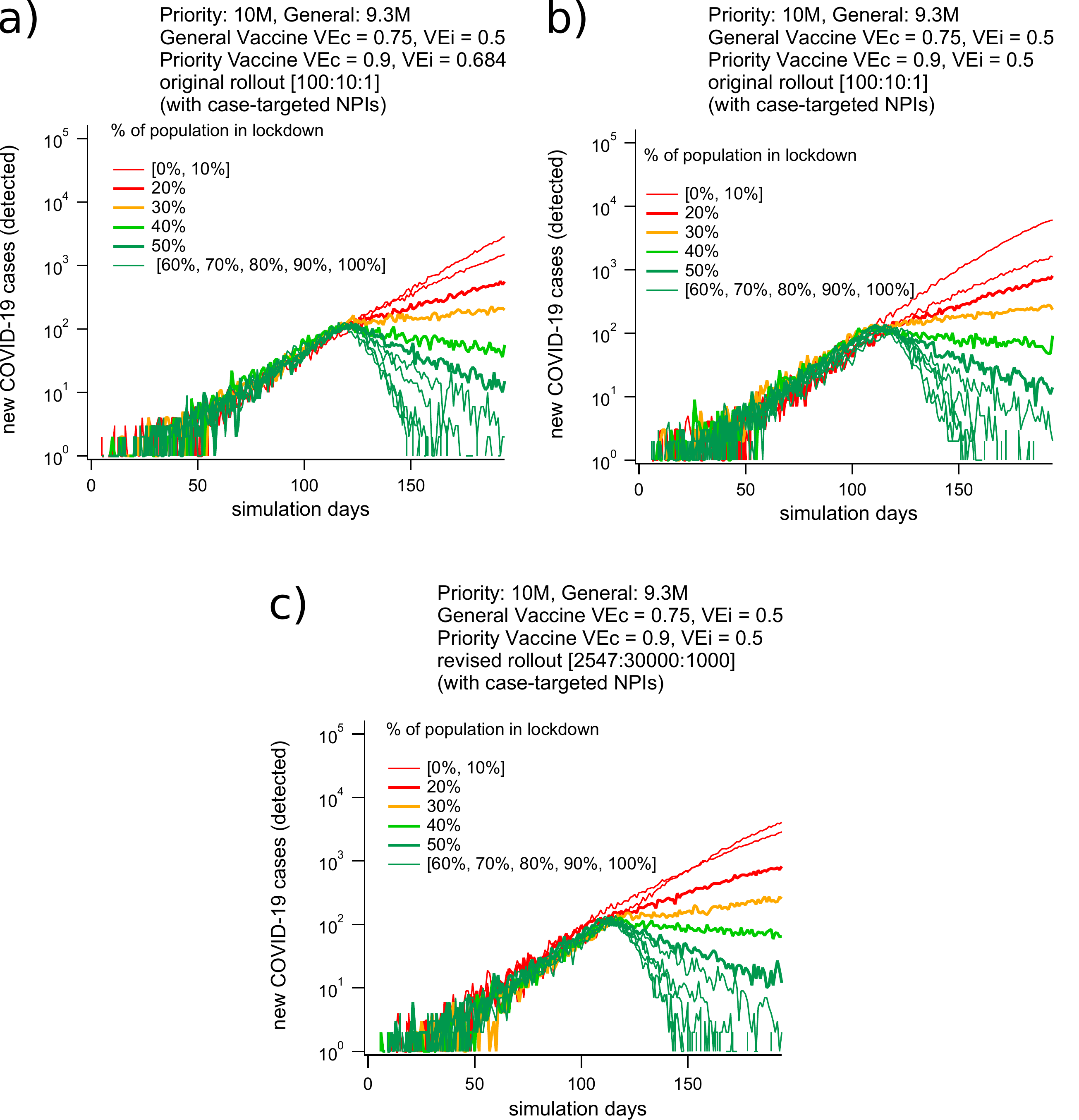}
    \caption{Within realistic parameter ranges, the lockdown compliance thresholds for elimination are not sensitive to vaccine efficacy assumptions or vaccine priority for different age groups. Representative incidence trajectories for different lockdown compliance rates demonstrate an elimination threshold between 30\% and 40\% for three alternative vaccine rollout scenarios. The trajectories in (a) correspond to increasing the three efficacy components for the general vaccine from 0.368 to 0.5, to approximate estimates from recent population-scale trial data. The trajectories in (b) further modify the efficacy of the priority vaccine against infectiousness from 0.684 to 0.5, reducing it in line with recent estimates. In (c), an alternate priority schedule is tested, in which adults between the ages of 18 and 64 are prioritised over those aged 65+, to approximate recent changes to recommendations against the use of general vaccine, ChAdOx1 nCoV-19 (Oxford/AstraZeneca), in younger adults.}
    \label{fig:SX_sensitivity_trajectories}
\end{figure}

\begin{figure}[h]
    \centering
    \includegraphics[width=0.8\textwidth]{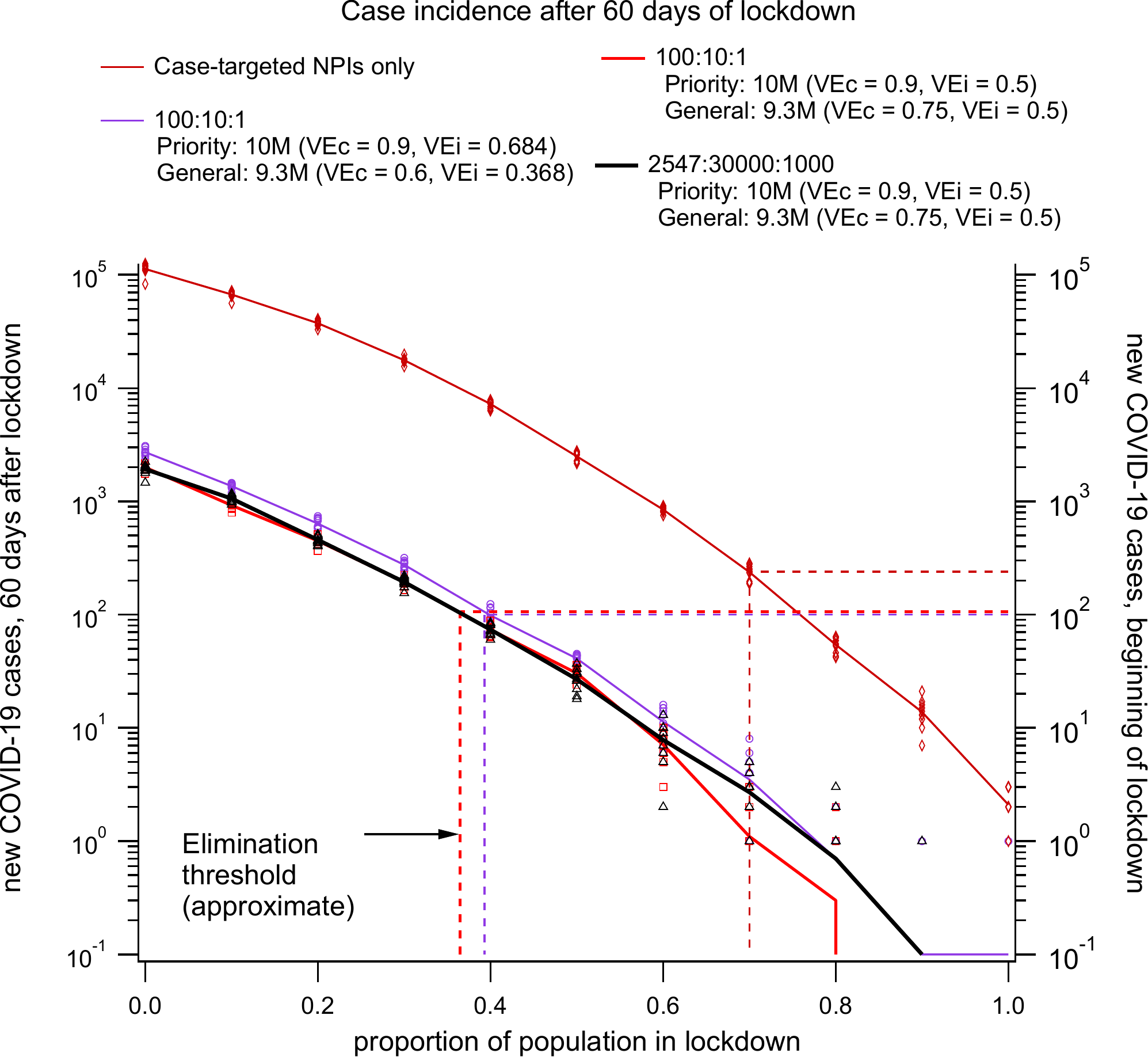}
    \caption{The intensity of lockdown required for gradual elimination of the virus is not sensitive to revised vaccine efficacy estimates or to a revised priority schedule. Solid lines connect ensemble averages of case incidence 60 days into the lockdown period for each scenario (left y-axis) while the values recorded from each individual simulation are shown as symbols. Each horizontal dashed line corresponds to the average incidence at the onset of lockdown (right y axis) for the vaccination scenario labelled with the same colour. The vertical dashed lines correspond to the approximate proportion of the population in lockdown required for case incidence to decrease, in each vaccination scenario. We omitted black dashed lines corresponding to the revised scenario with updated vaccine efficacy because they overlap closely with the red dashed lines corresponding to the original priority schedule with updated vaccine efficacy.}
    \label{fig:SX_sensitivity_SD_thresh}
\end{figure}

\FloatBarrier
\newpage

\subsection*{Epidemic severity as a function of vaccine efficacy and coverage}

Here we provide full ABM results for epidemic severity as measured by initial incidence growth rates, peak prevalence levels and the timing of the prevalence peaks from the start of the epidemic as functions of $\VEi$, $\VEs$, $\VEd$, and coverage. Incidence growth rate is computed from the first detected case to the time at which cumulative cases exceed 2000, or the end of the simulation after 194 days.


\begin{table}[ht]
    \centering
\includegraphics[width=\textwidth]{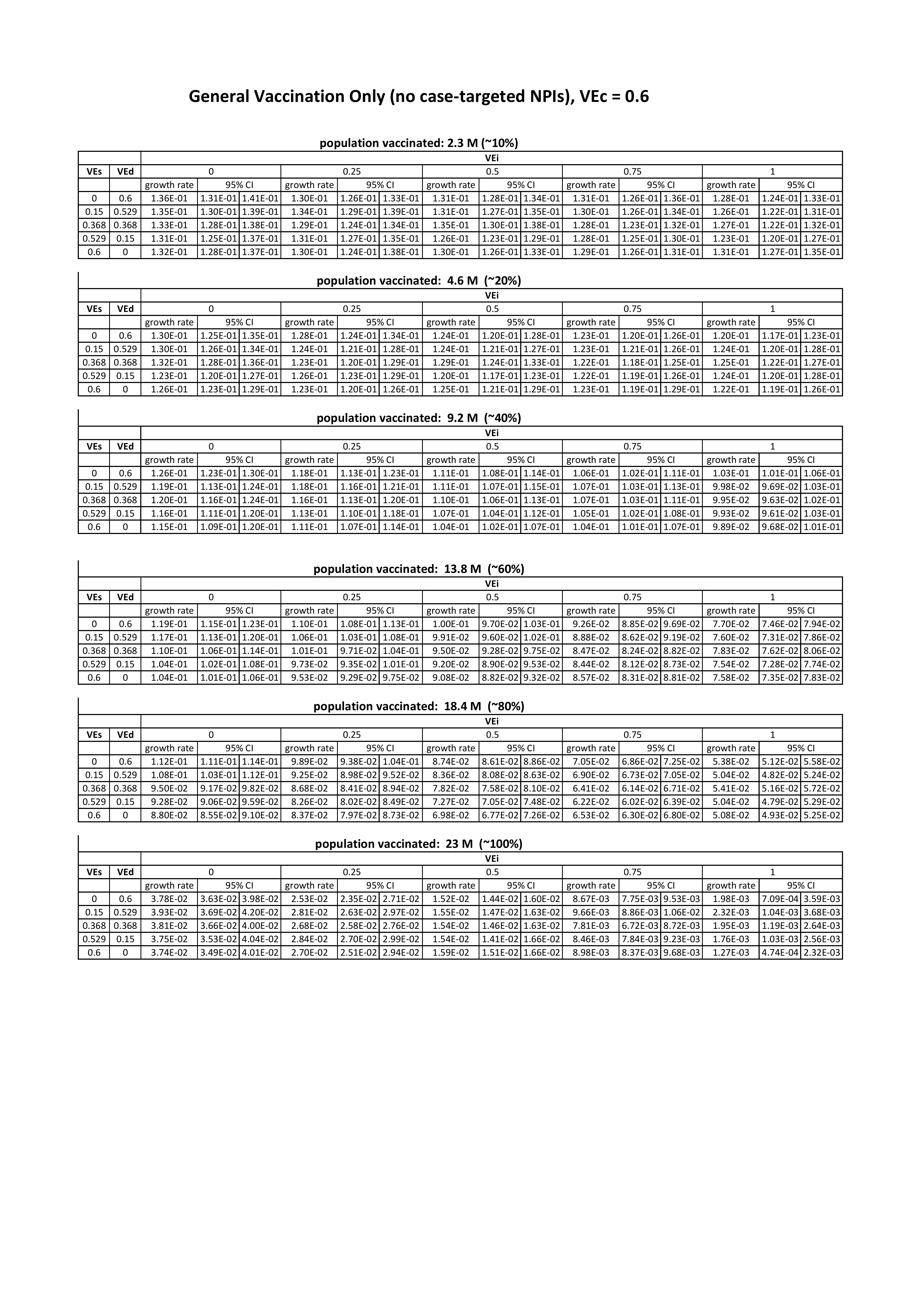}
\vspace*{-60mm}
    \caption{Mean incidence growth rate values and 95\% ensemble CIs produced by the ABM for various combinations of vaccine efficacy parameters and coverage levels, assuming a clinical efficacy of $\VEc = 0.6$ and no case-targeted NPIs (n = 10 instances per scenario).}
    \label{tab:general_no_ctnpi_gr_first}
\end{table}

\begin{table}[ht]
    \centering
\includegraphics[width=\textwidth]{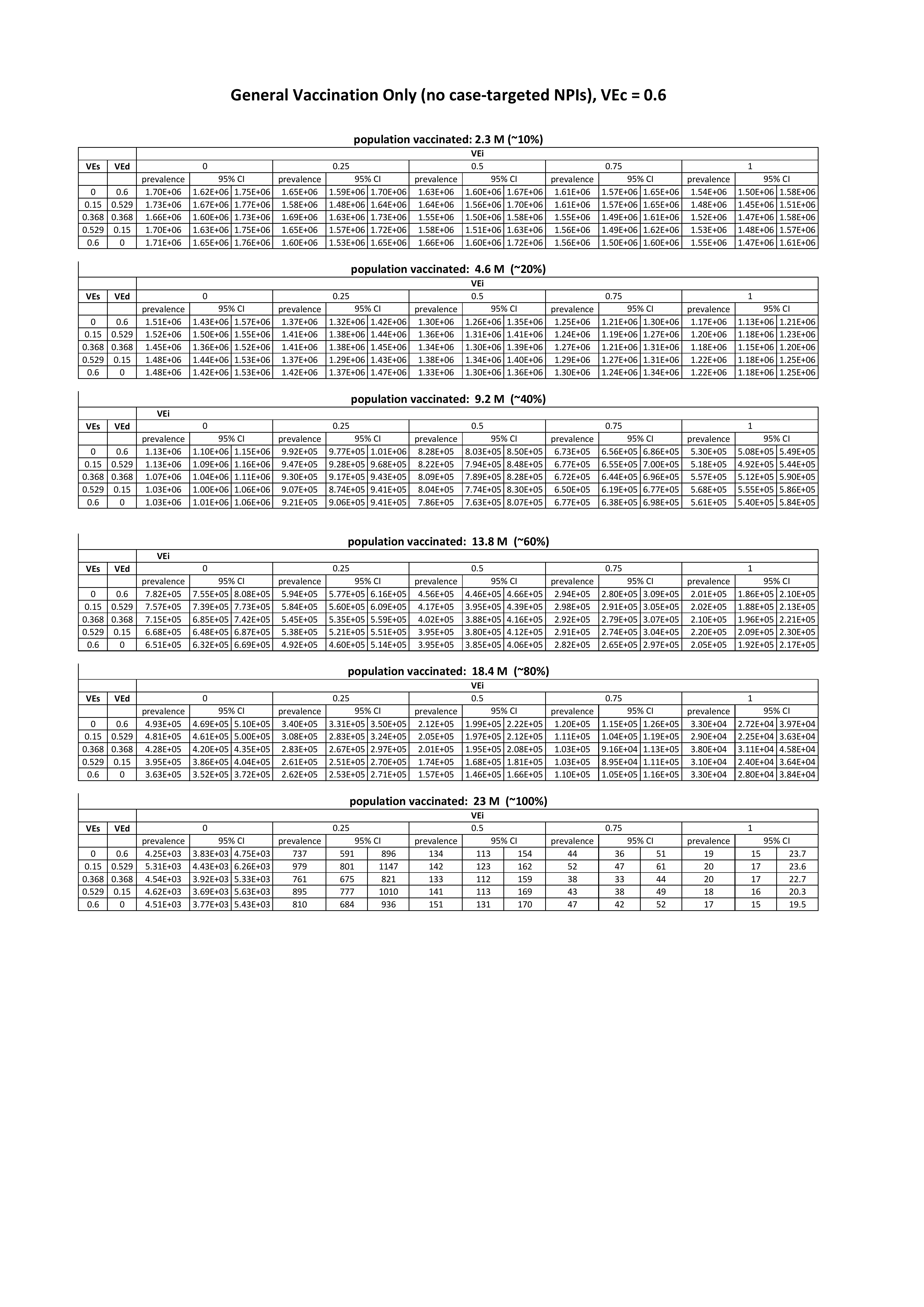}
\vspace*{-60mm}
    \caption{Mean peak prevalence values and 95\% ensemble bootstrap CIs produced by the ABM for various combinations of vaccine efficacy parameters and coverage levels, assuming a clinical efficacy of $\VEc = 0.6$ and no case-targeted NPIs (n = 10 instances per scenario).}
    \label{tab:general_no_ctnpi}
\end{table}

\begin{table}[ht]
    \centering
\includegraphics[width=\textwidth]{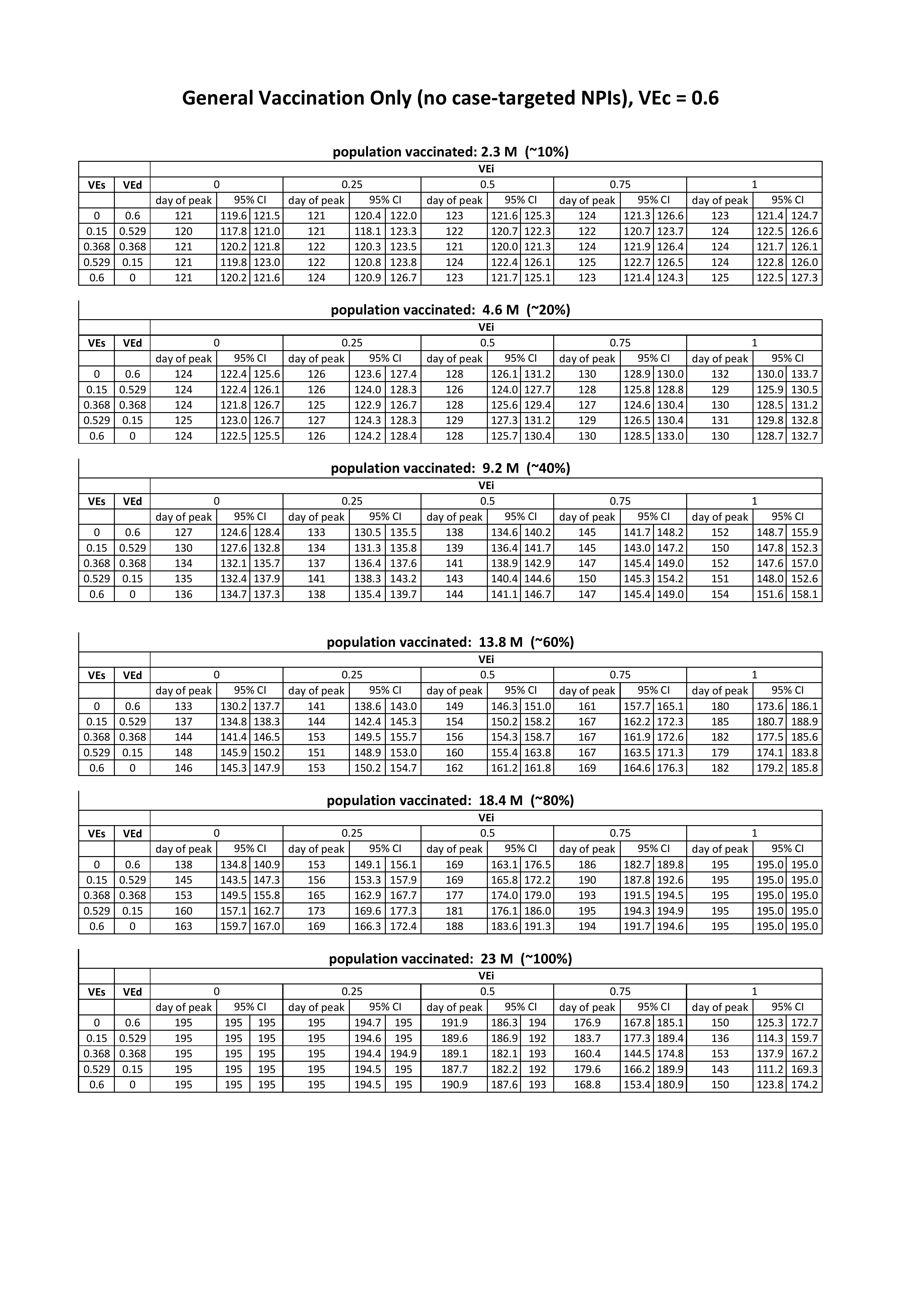}
\vspace*{-40mm}
    \caption{Prevalence peak times (means and 95\% ensemble bootstrap CIs) produced by the ABM for various combinations of vaccine efficacy parameters and coverage levels, assuming a clinical efficacy of $\VEc = 0.6$ and no case-targeted NPIs (n = 10 instances per scenario).}
    \label{tab:general_no_ctnpi_timing}
\end{table}


\begin{table}[ht]
    \centering
\includegraphics[width=\textwidth]{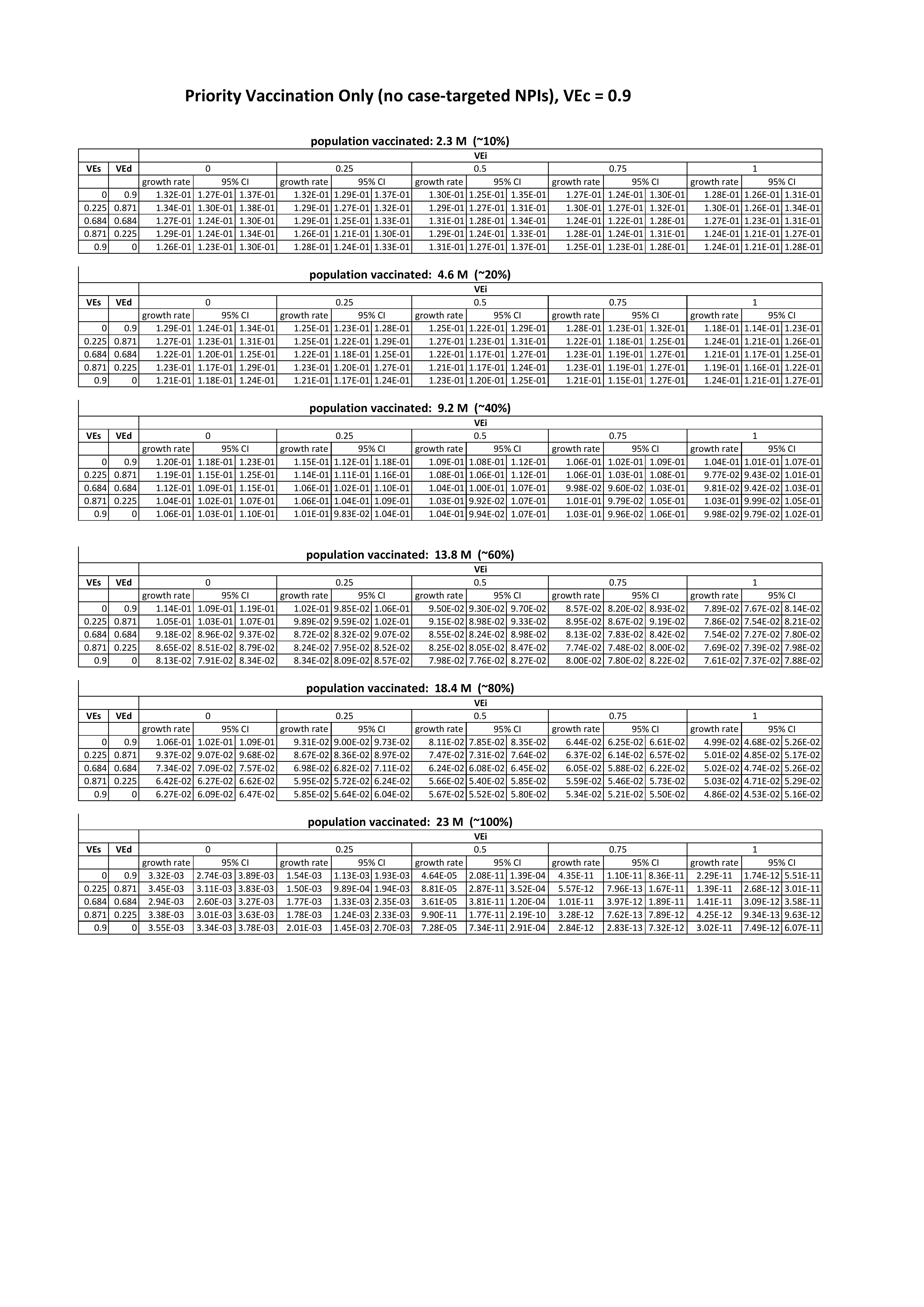}
\vspace*{-60mm}
    \caption{Mean incidence growth rate values and 95\% ensemble bootstrap CIs produced by the ABM for various combinations of vaccine efficacy parameters and coverage levels, assuming a clinical efficacy of $\VEc = 0.9$ and no case-targeted NPIs (n = 10 instances per scenario).}
    \label{tab:priority_no_ctnpi_gr}
\end{table}

\begin{table}[ht]
    \centering
\includegraphics[width=\textwidth]{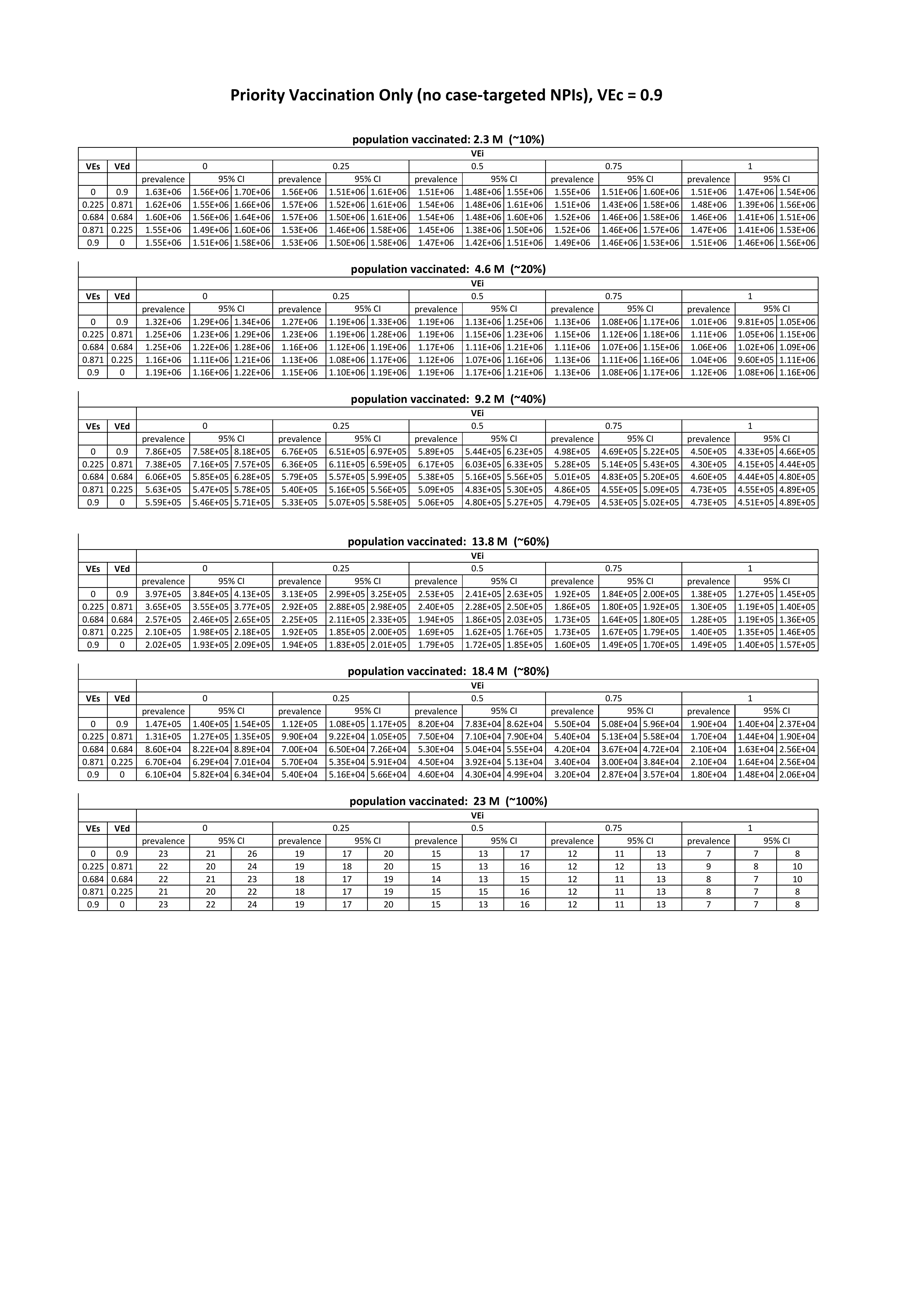}
\vspace*{-60mm}
    \caption{Mean peak prevalence values and 95\% ensemble bootstrap CIs produced by the ABM for various combinations of vaccine efficacy parameters and coverage levels, assuming a clinical efficacy of $\VEc = 0.9$ and no case-targeted NPIs (n = 10 instances per scenario).}
    \label{tab:priority_no_ctnpi}
\end{table}

\begin{table}[ht]
    \centering
\includegraphics[width=\textwidth]{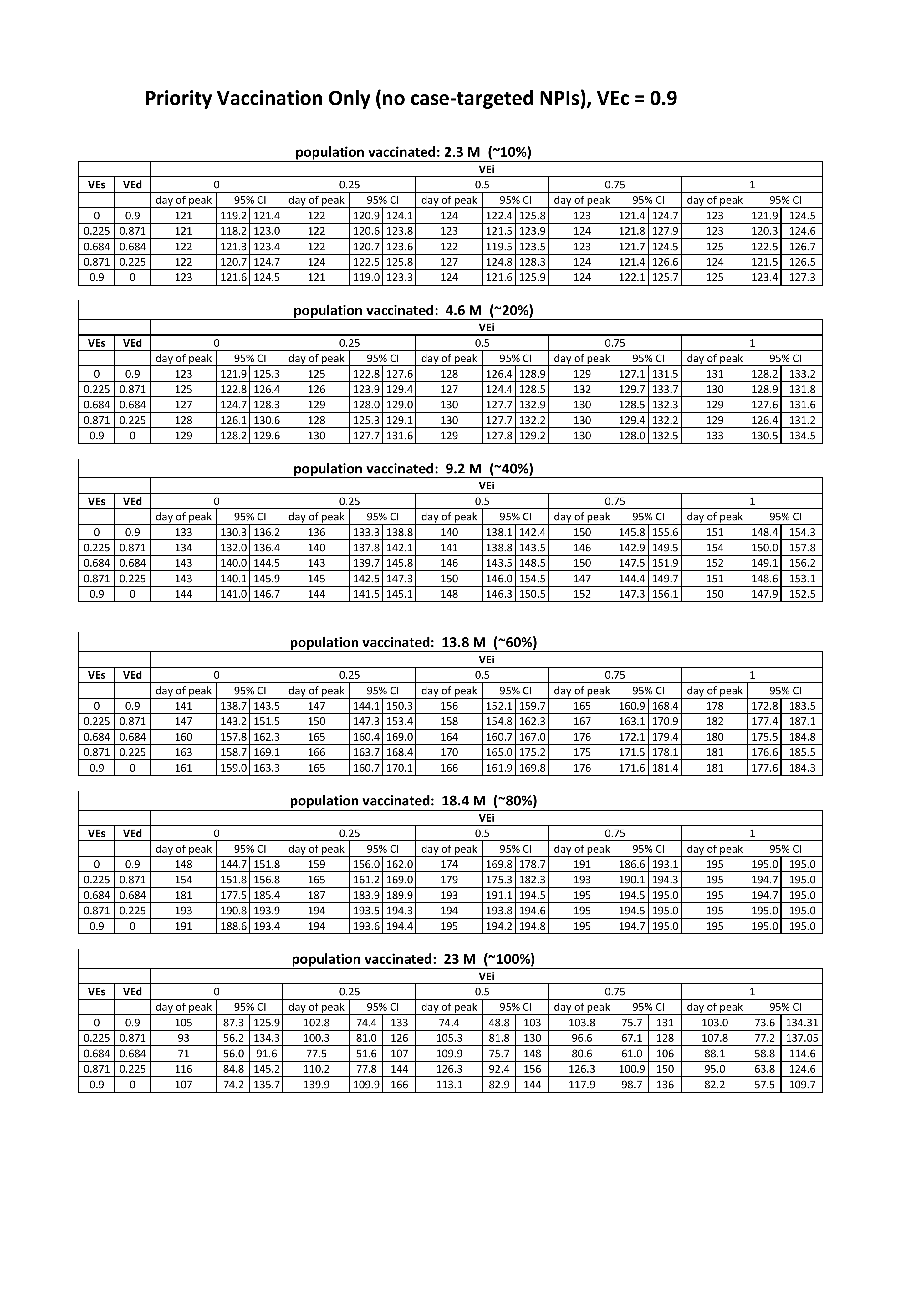}
\vspace*{-40mm}
    \caption{Prevalence peak times (means and 95\% ensemble bootstrap CIs) produced by the ABM for various combinations of vaccine efficacy parameters and coverage levels, assuming a clinical efficacy of $\VEc = 0.9$ and no case-targeted NPIs (n = 10 instances per scenario).}
    \label{tab:priority_no_ctnpi_timing}
\end{table}


\begin{table}[ht]
    \centering
\includegraphics[width=\textwidth]{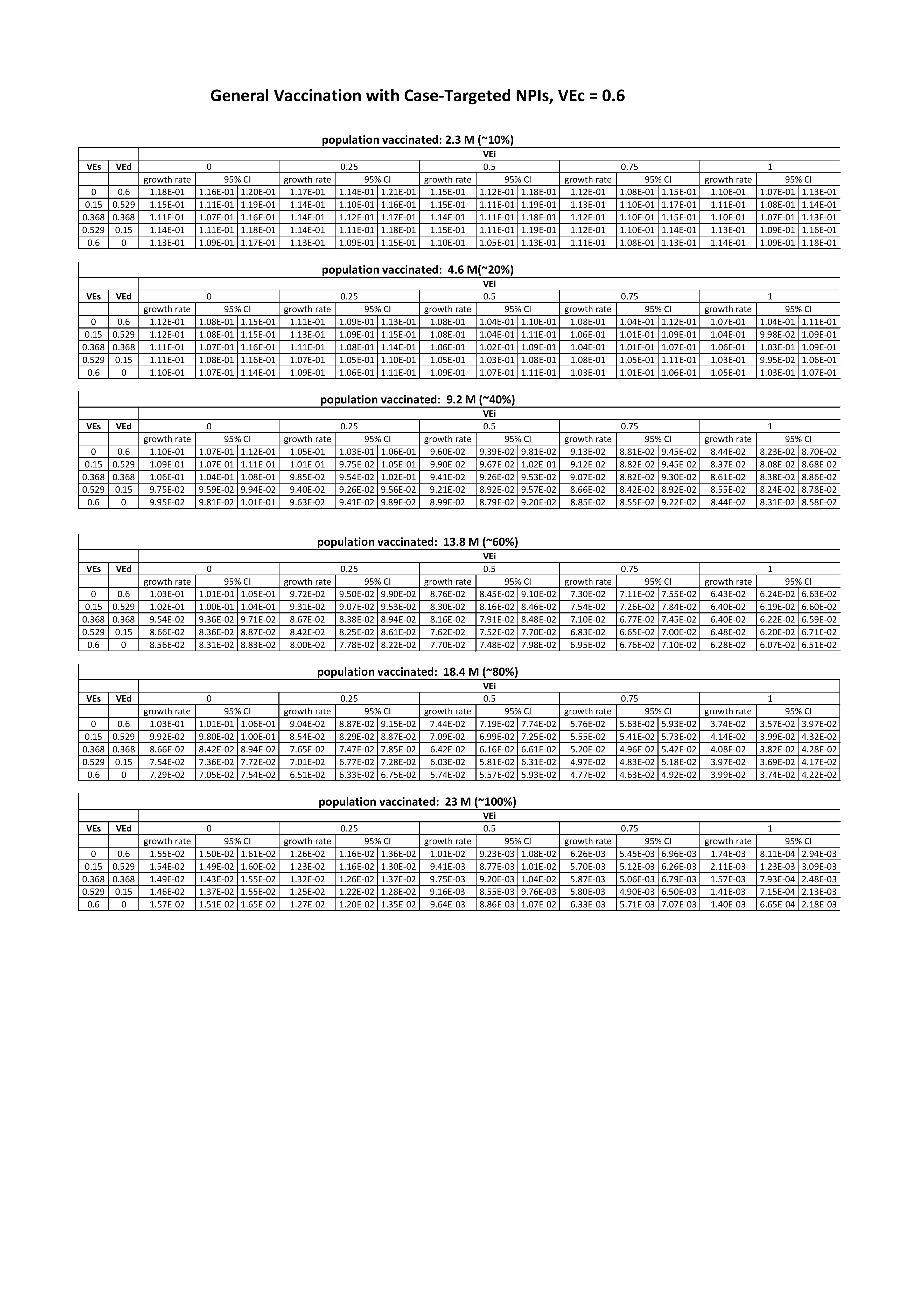}
\vspace*{-60mm}
    \caption{Mean incidence growth rate values and 95\% ensemble bootstrap CIs produced by the ABM for various combinations of vaccine efficacy parameters and coverage levels, assuming a clinical efficacy of $\VEc = 0.6$ and active case-targeted NPIs consisting of case isolation, home-quarantine of household contacts of detected cases, and international travel restrictions (n = 10 instances per scenario).}
    \label{tab:general_ctnpi_gr}
\end{table}

\begin{table}[ht]
    \centering
\includegraphics[width=\textwidth]{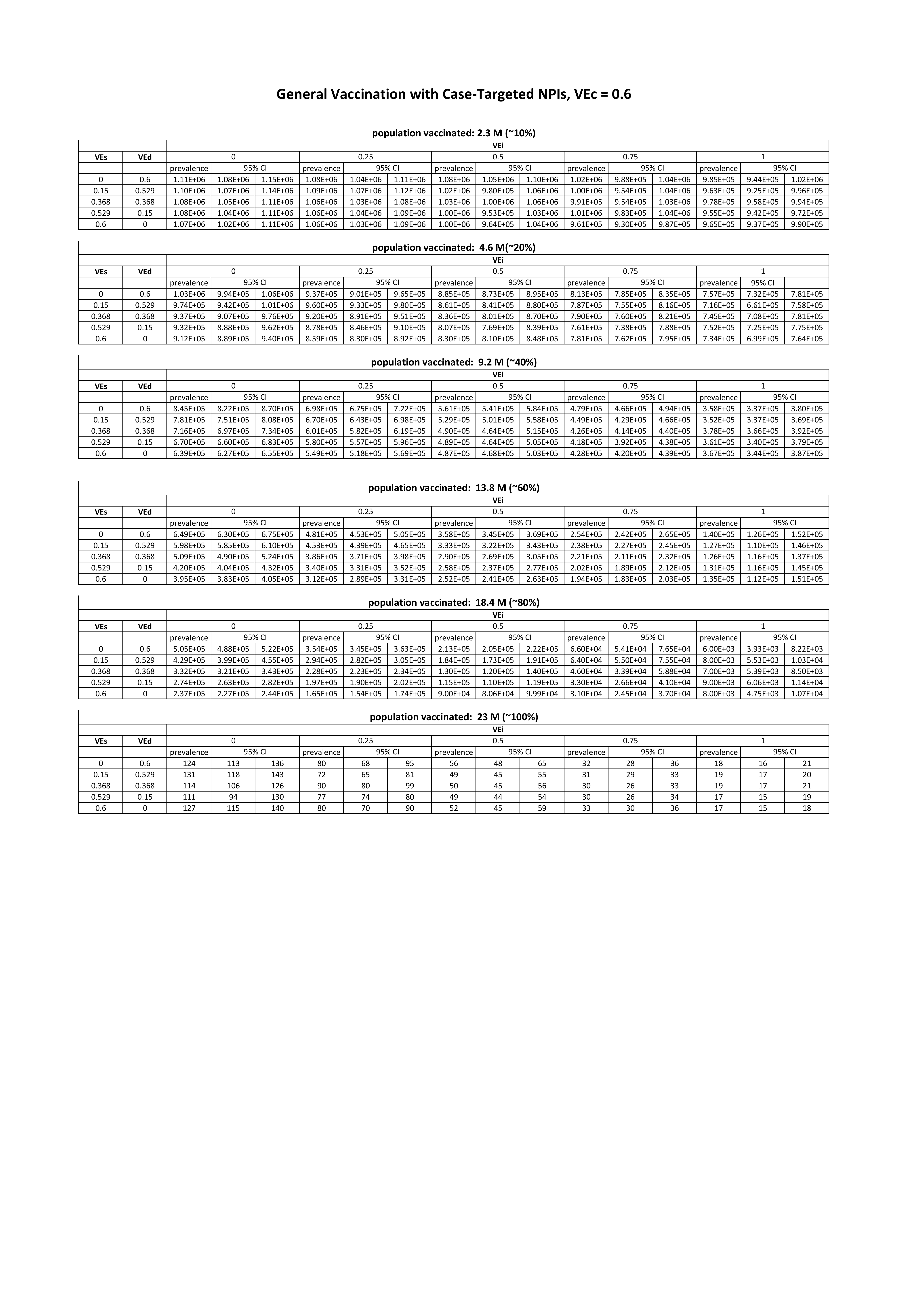}
\vspace*{-80mm}
    \caption{Mean peak prevalence values and 95\% ensemble bootstrap CIs produced by the ABM for various combinations of vaccine efficacy parameters and coverage levels, assuming a clinical efficacy of $\VEc = 0.6$ and active case-targeted NPIs consisting of case isolation, home-quarantine of household contacts of detected cases, and international travel restrictions (n = 10 instances per scenario).}
    \label{tab:general_ctnpi}
\end{table}

\begin{table}[ht]
    \centering
    \includegraphics[width=\textwidth]{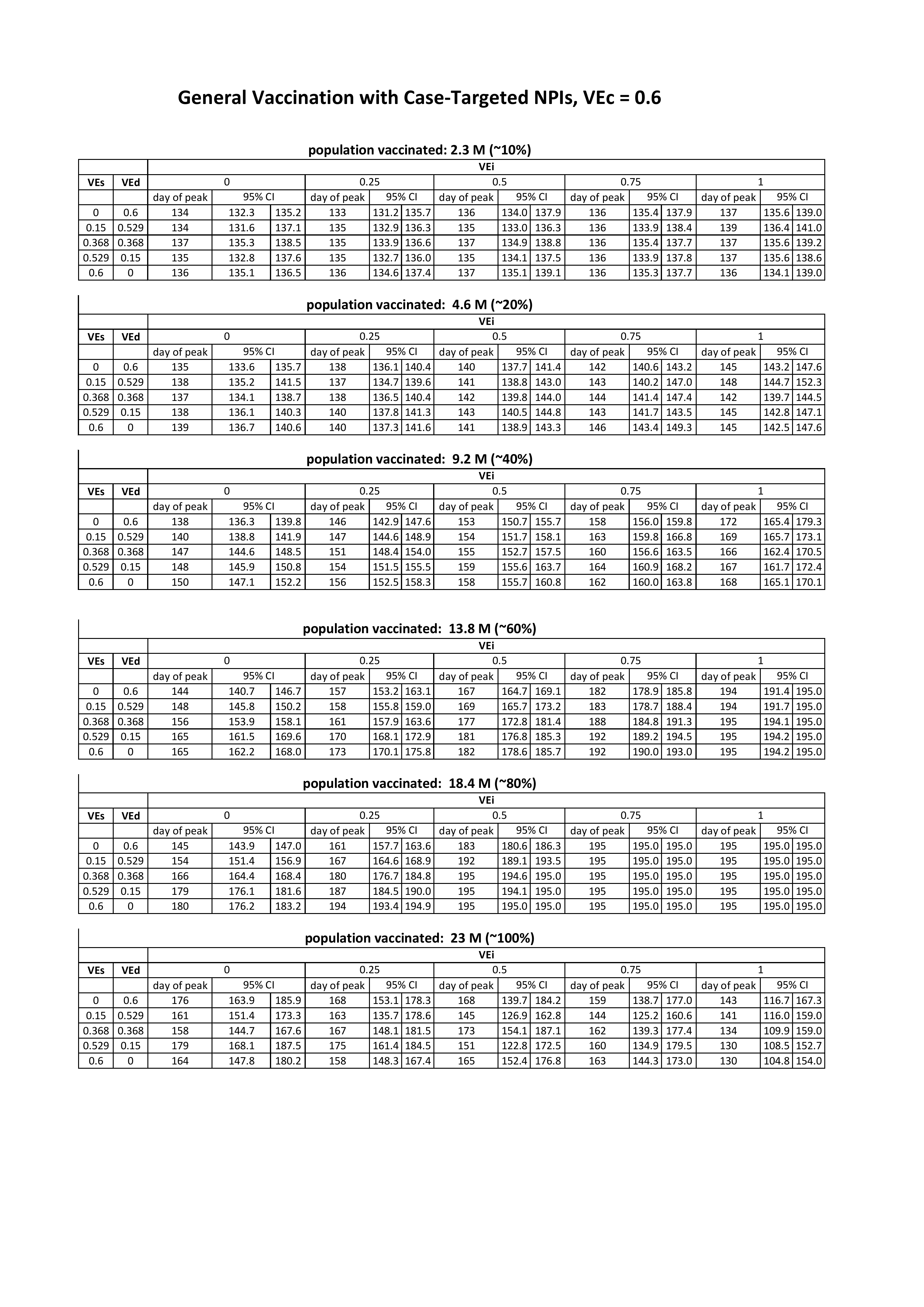}
		\vspace*{-40mm}
    \caption{Prevalence peak times (means and 95\% ensemble bootstrap CIs) produced by the ABM for various combinations of vaccine efficacy parameters and coverage levels, assuming a clinical efficacy of $\VEc = 0.6$ and active case-targeted NPIs consisting of case isolation, home-quarantine of household contacts of detected cases, and international travel restrictions (n = 10 instances per scenario).}
    \label{tab:general_ctnpi_timing}
\end{table}


\begin{table}[ht]
    \centering
    \includegraphics[width=\textwidth]{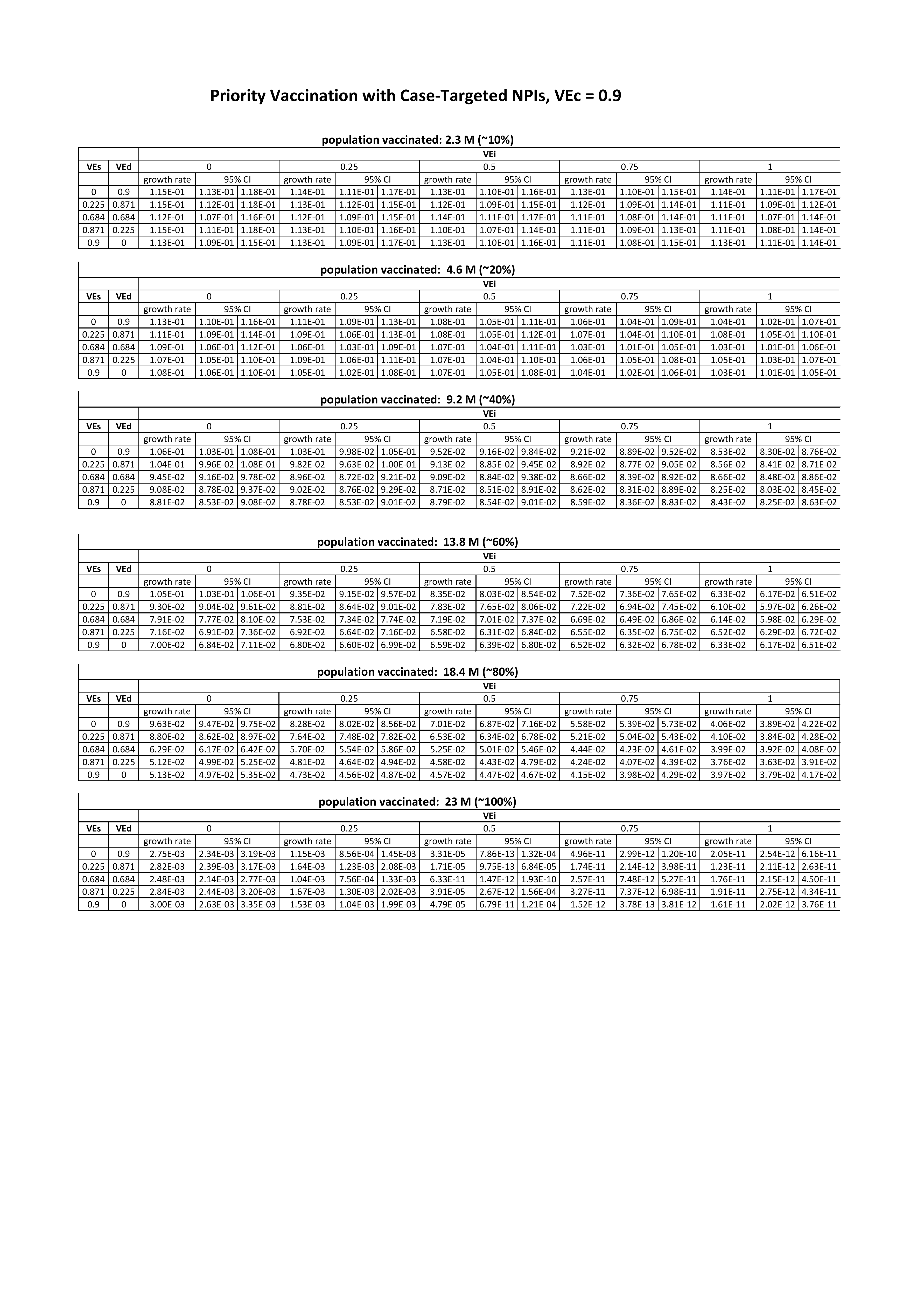}
		\vspace*{-60mm}
    \caption{Mean incidence growth rate and 95\% ensemble bootstrap CIs produced by the ABM for various combinations of vaccine efficacy parameters and coverage levels, assuming a clinical efficacy of $\VEc = 0.9$ and active case-targeted NPIs consisting of case isolation, home-quarantine of household contacts of detected cases, and international travel restrictions (n = 10 instances per scenario).}
    \label{tab:priority_ctnpi_gr}
\end{table}

\begin{table}[ht]
    \centering
    \includegraphics[width=\textwidth]{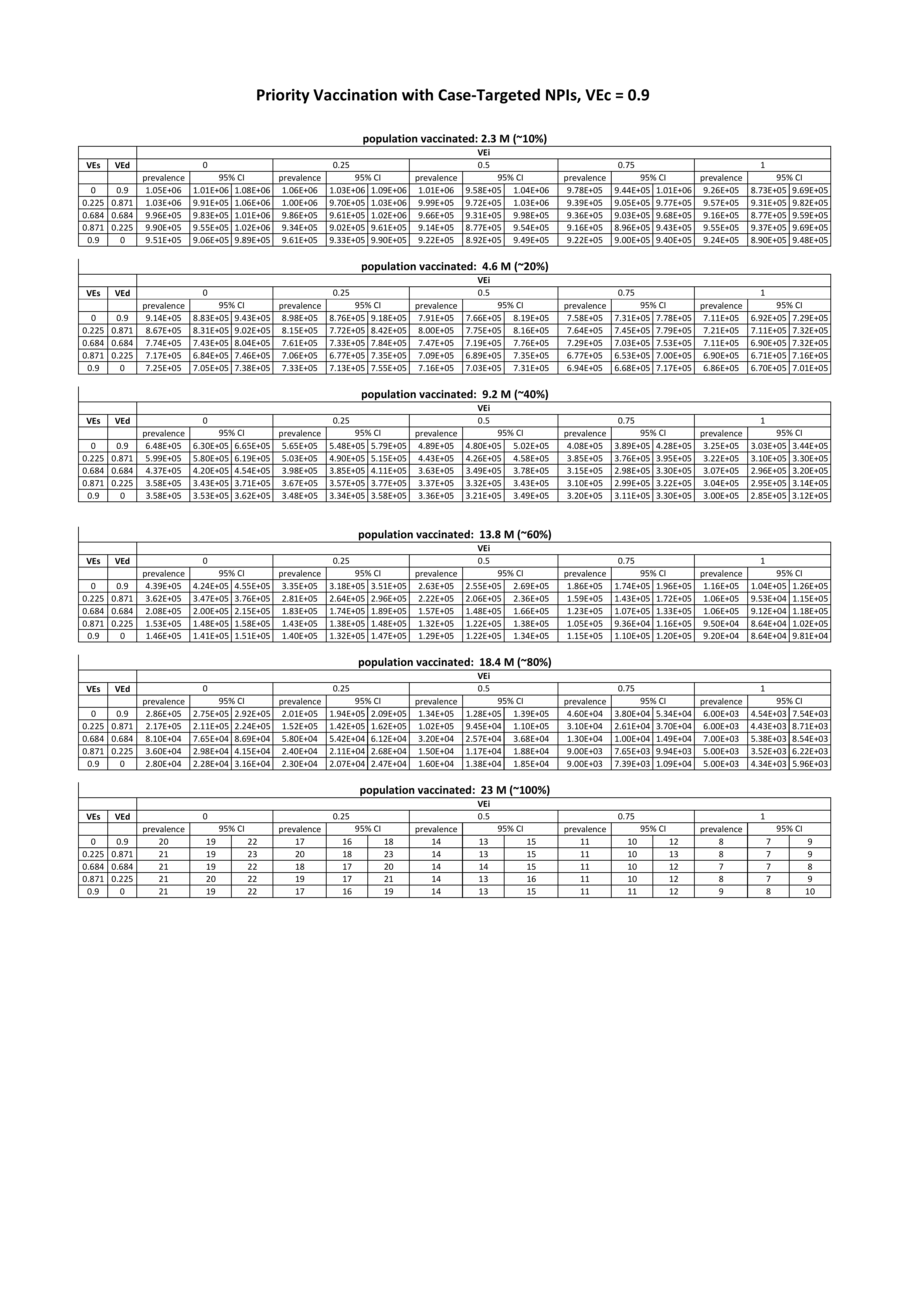}
		\vspace*{-60mm}
    \caption{Mean peak prevalence values and 95\% ensemble bootstrap CIs produced by the ABM for various combinations of vaccine efficacy parameters and coverage levels, assuming a clinical efficacy of $\VEc = 0.9$ and active case-targeted NPIs consisting of case isolation, home-quarantine of household contacts of detected cases, and international travel restrictions (n = 10 instances per scenario).}
    \label{tab:priority_ctnpi}
\end{table}

\begin{table}[ht]
    \centering
    \includegraphics[width=\textwidth]{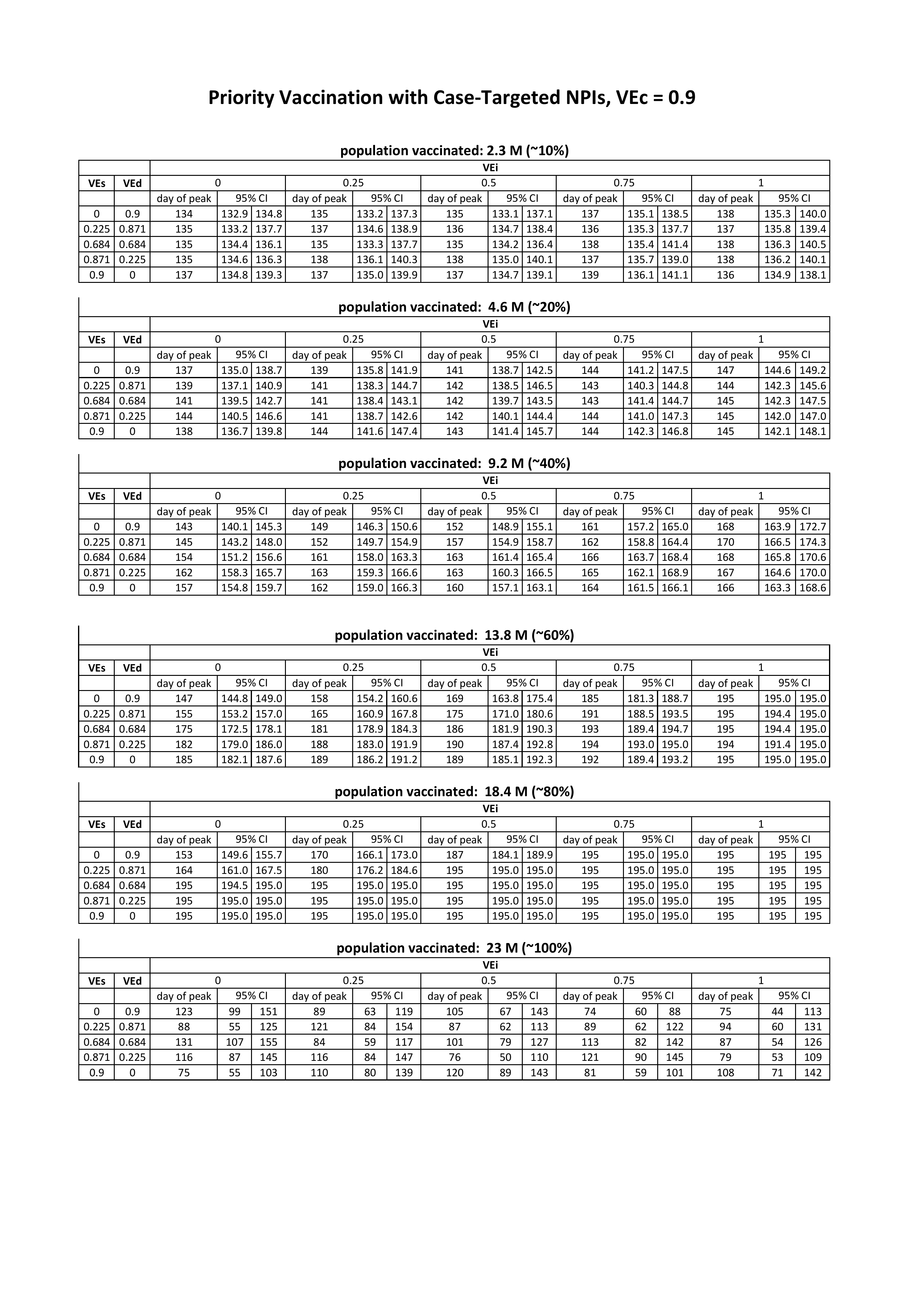}
		\vspace*{-40mm}
    \caption{Prevalence peak times (means and 95\% ensemble bootstrap CIs) produced by the ABM for various combinations of vaccine efficacy parameters and coverage levels, assuming a clinical efficacy of $\VEc = 0.9$ and active case-targeted NPIs consisting of case isolation, home-quarantine of household contacts of detected cases, and international travel restrictions (n = 10 instances per scenario).}
    \label{tab:priority_ctnpi_timing_last}
\end{table}

\clearpage


\end{document}